\documentclass[journal=jctcce, layout=traditional]{achemso}%

\usepackage{chemformula} 
\usepackage[T1]{fontenc} 

\usepackage{amsmath,color}
\usepackage{amssymb}
\usepackage{amsfonts}
\usepackage{amsthm}
\usepackage{mathtools}

\usepackage[title]{appendix}

\usepackage{algorithm}
\usepackage{algpseudocode}
\usepackage{dsfont}

\DeclarePairedDelimiter\bra{\langle}{\rvert}
\DeclarePairedDelimiter\ket{\lvert}{\rangle}
\DeclarePairedDelimiterX\braket[2]{\langle}{\rangle}{#1 \delimsize\vert #2}

\newcommand{\vJ}{\mathbf{J}}
\newcommand{\vP}{{\boldsymbol{\mathcal{P}}}}
\newcommand{\vPtot}{{\mathbf{P}}}

\newcommand{\vC}{\mathbf{C}}
\newcommand{\vE}{\mathbf{E}}
\newcommand{\vEin}{\mathbf{E}_{\text{in}}}
\newcommand{\vEscatt}{\mathbf{E}_{\text{scatt}}}
\newcommand{\vB}{\mathbf{B}}
\newcommand{\vD}{\mathbf{D}}
\newcommand{\ve}{\mathbf{e}}
\newcommand{\vnabla}{\boldsymbol{\nabla}}
\newcommand{\vdelta}{\boldsymbol{\delta}}
\newcommand{\vmu}{\boldsymbol{\mu}}
\newcommand{\hmu}{\hat{\boldsymbol{\mu}}}
\newcommand{\vH}{\mathbf{H}}

\newcommand{\vR}{\mathbf{r}}

\newcommand{\vxi}{\boldsymbol{\xi}}
\newcommand{\kFGR}{k_{\text{FGR}}}
\newcommand{\kEh}{k_{\text{Eh}}}
\newcommand{\gEh}{\gamma_{\text{Eh}}}
\newcommand{\gR}{\gamma_{\text{R}}}
\newcommand{\kR}{k_{\text{R}}}
\newcommand{\he}{\mathbf{e}}
\newcommand{\hH}{\hat{H}}

\newcommand{\hE}{\hat{\mathbf{E}}}
\newcommand{\hB}{\hat{\mathbf{B}}}

\newcommand{\hsigma}{\hat{\sigma}}

\newcommand{\hP}{\hat{\boldsymbol{\mathcal{P}}}}
\newcommand{\hrho}{\hat{\rho}}

\newcommand{\Ut}{U_{\text{tot}}}
\newcommand{\Us}{U_{\text{s}}}
\newcommand{\UEM}{U_{\text{EM}}}
\newcommand{\UR}{U_{\text{R}}}

\newcommand{\TkR}{ {\hat{\mathcal{T}}_{1\leftarrow 2}[k_{\text{R}}] } }

\newcommand{\tr}[1]{\text{Tr}\left(#1\right)}
\newcommand{\Lrho}[1]{\mathcal{L}_{\text{#1}}[\hat{\rho}]}
\newcommand{\avg}[1]{\left\langle #1\right\rangle}

	\title{ A Comparison of Different Classical, Semiclassical and Quantum Treatments of Light--Matter Interactions: Understanding Energy Conservation}%
	
	\author{Tao E. Li}%
	\email{taoli@sas.upenn.edu}
	\affiliation{Department of Chemistry, University of Pennsylvania, Philadelphia, Pennsylvania 19104, USA}

	\author{Hsing-Ta Chen}
	\affiliation{Department of Chemistry, University of Pennsylvania, Philadelphia, Pennsylvania 19104, USA}
	
	\author{Joseph E. Subotnik}
	\email{subotnik@sas.upenn.edu}
	\affiliation{Department of Chemistry, University of Pennsylvania, Philadelphia, Pennsylvania 19104, USA}

\begin{document}
	
	\begin{abstract}
		The optical response of an electronic two-level system (TLS) coupled to an incident continuous wave (cw) electromagnetic (EM) field is simulated explicitly in one dimension by the following five approaches:  (i) the coupled Maxwell--Bloch equations, (ii) the optical Bloch equation (OBE), (iii) Ehrenfest dynamics, (iv) the Ehrenfest+R approach and (v) classical dielectric theory (CDT). Our findings are as follows: (i) standard Ehrenfest dynamics predict the correct optical signals only in the linear response regime where vacuum fluctuations are not important; (ii) both the coupled Maxwell--Bloch equations and CDT predict incorrect features for the optical signals in the linear response regime due to a double-counting of self-interaction;  (iii) by exactly balancing the effects of self-interaction versus the effects of quantum fluctuations (and insisting on energy conservation), the Ehrenfest+R approach generates the correct optical signals in the linear regime and slightly beyond, yielding, e.g., the correct ratio between the coherent and incoherent scattering EM fields. As such, Ehrenfest+R dynamics agree with dynamics from the quantum OBE, but whereas the latter is easily applicable only for a single TLS in vacuum, the former should be applicable to large systems in environments with arbitrary dielectrics. Thus, this benchmark study suggests that the Ehrenfest+R approach may be very advantageous for simulating light--matter interactions semiclassically.
	\end{abstract}

	\maketitle
\begin{tocentry} 
	\includegraphics[width=6.3cm]{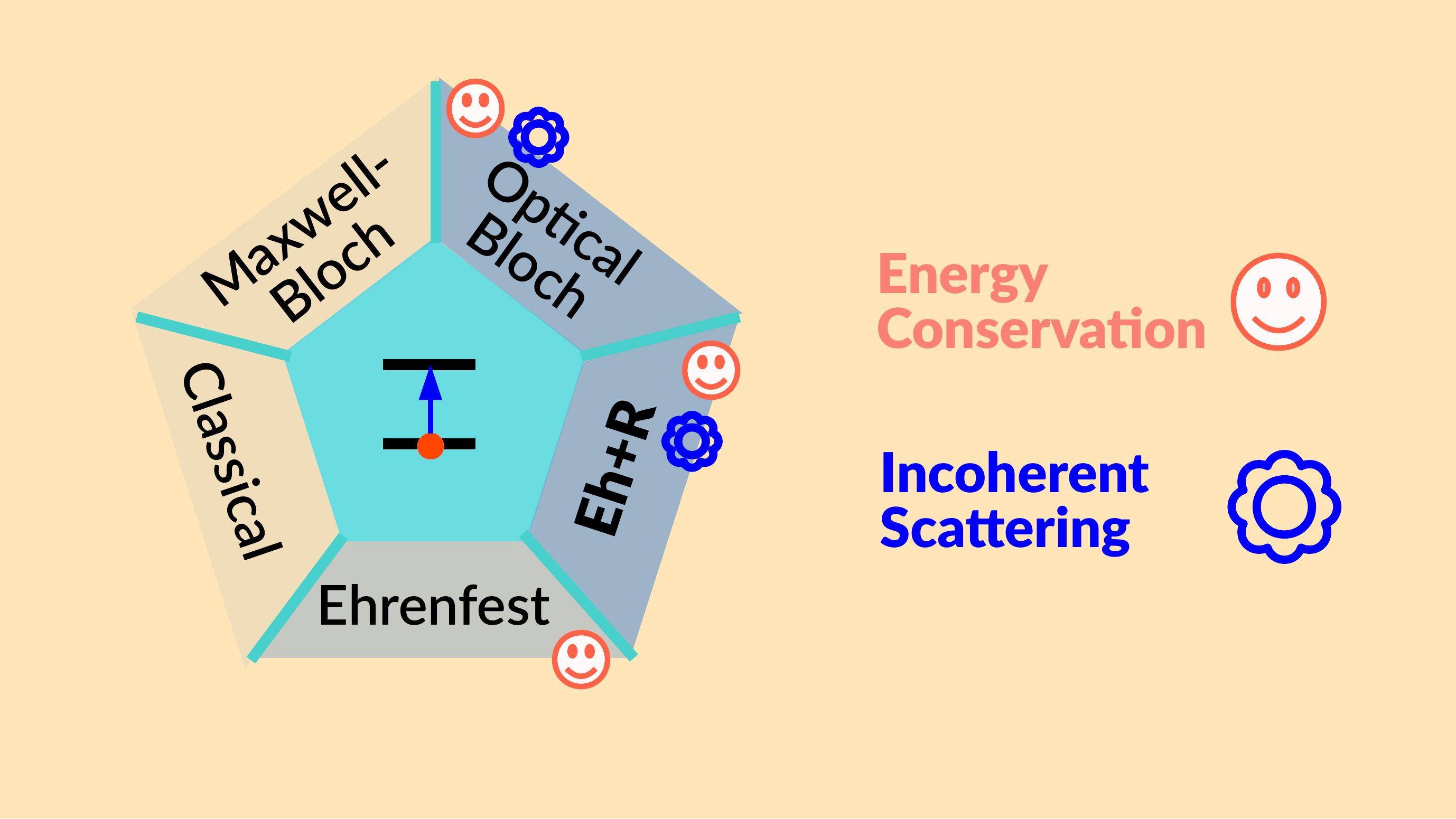}
\end{tocentry}

\section{Introduction}\label{sec:intro}
The nature of light--matter interactions is a never-ending source of stimulation in both experimental and theoretical science. To theoretically study light--matter interactions at the atomic and molecular level, non-relativistic quantum electrodynamics (QED)\cite{AKBARSALAM2010} is the ultimate theory: the matter side obeys the Schr\"odinger equation (or the quantum Liouville equation) and the radiation field is quantized as photons. And for weak light--matter interactions, perturbative approximations of QED are both practical and  adequate for describing most experimental findings. However, to understand recent experiments that involve strong light--matter interactions\cite{Gibbs2011,Lodahl2015,Ribeiro2018,Sukharev2017,Torma2015,Vasa2018},  one must go beyond the perturbative limit of QED. Furthermore, if we cannot use linear response theory, then both matter and photons must be simulated explicitly and on the same footing. Moreover, because photons are described in a vast infinite-dimensional Hilbert space, a rigorous QED treatment becomes a computational nightmare for calculations applicable to most modern work in nanoscience.

To reduce the computational cost of QED, one plausible simplification is to work in a truncated space of photons, i.e., to consider a subspace with either only a few photon modes or a few photon quanta per mode. In the framework of this simplification, one approach is to invoke Floquet theory\cite{Grifoni1998,Hausinger2010}; another approach is to construct a dressed state representation\cite{Nitzan2006}. Both techniques can accurately describe strong light--matter interactions. Unfortunately, however, these techniques are usually applicable only when modeling either systems exposed to cw light or systems encapsulated in model three-dimensional (3D) cavities.

For large scale simulations of material systems in arbitrary EM fields and not necessarily in a full-3D cavity, a more promising ansatz would appear to be semiclassical electrodynamics\cite{Milonni1976, Sukharev2011,Heller2018}, according to which the matter side is described quantum-mechanically and the radiative electromagnetic (EM) fields are treated classically. More precisely, one replaces the EM field operators (acting in an infinite-dimensional Hilbert space) by c-numbers and propagates the dynamics of the EM field using the classical Maxwell's equations. Compared with QED calculations, semiclassical approaches are computationally much more affordable, and one can still preserve more of the quantum nature of light--matter interactions than standard classical electrodynamics (where the matter side is described as a dielectric constant\cite{Griffiths1999}). Thus, semiclassical electrodynamics would appear to be a natural compromise between speed and accuracy.

Now, although semiclassical approaches have many appealing qualities, 
one inevitable question arises: can one actually capture any true quantum effects of the EM field using such techniques? For example, because of the nature of the quantum radiation field, an electron in an excited state can automatically decay to its ground state even when no external field is applied, a phenomenon known as spontaneous emission. 
As has been argued by many (including Cohen-Tannoudji)\cite{Senitzky1973,Milonni1975,Dalibard1982,Dalibard1984}, this spontaneous process can in fact be dissected into two well-defined sub-processes:
(i) first, \textit{vacuum fluctuations} that arise from the zero-point energy of the EM field; (ii) second, emission carried by electronic \textit{self-interaction} as the scattered EM field induced by the electron reacts back on the electron itself.
While the latter has a classical analog (e.g., the Abraham-Lorentz equation\cite{Jackson1999,Jimenez1987}),  the former is a purely quantum effect. Can spontaneous emission be captured by semiclassical approaches (i.e., Ehrenfest dynamics\cite{Mukamel1999,Li2018Spontaneous,Li2018Tradeoff})? In fact, it is well-known that, according to Ehrenfest dynamics (or neoclassical electrodynamics\cite{Barut1980}), only self-interaction is taken into account in the equation of motion for the electron: vacuum fluctuations are not included. As a result, an electron is stabilized in the excited state according to Ehrenfest dynamics, thus  violating many experimental observations of florescence. A lack of spontaneous emission is one failure of semiclassical electrodynamics theory, and one would like to go beyond Ehrenfest when simulating large systems where spontaneous emission cannot be ignored.

\subsection{Beyond Ehrenfest dynamics}
Perhaps the most straightforward way to correct Ehrenfest dynamics so as to include spontaneous emission is to add a linear, "hard" dissipative term on top of the Liouville equation for the matter side and thus force the electron to decay without concern for the dynamics of the EM fields. 
This approach makes sense  since the overall spontaneous decay rate in vacuum (as carried by self-interaction plus vacuum fluctuations) is a constant --- known as the Fermi's golden rule (FGR) rate --- regardless of the instantaneous electronic state.
By further assuming that the external field influences the electron in a classical way and does not alter the spontaneous decay rate of the electron, one arrives at the coupled Maxwell--Bloch equations\cite{Castin1995,Ziolkowski1995} for modeling light--matter interactions. According to a na\"ive implementation of the coupled Maxwell--Bloch equations, the EM field that one molecule feels is composed of both the incident externally imposed EM field plus the scattered field generated by the molecule itself. Given that Ehrenfest dynamics already incorporates self-interaction (as just discussed; see Refs. \citenum{Li2018Spontaneous,Chen2018Spontaneous}), it is then perhaps not surprising that one can show that a na\"ive Maxwell--Bloch scheme actually double-counts the electronic self-interaction\cite{Lopata2009-1}.

To avoid such nonphysical double-counting, many additional efforts have been made to exclude the self-interaction component of the EM field so that one need not modify the linear dissipative term within the Maxwell--Bloch scheme. A separation of the EM field into incident plus self-interacting components is possible because of the linearity of Maxwell's equation so that in principle we can insist that the molecule explicitly experiences only the incident field.\cite{Neuhauser2007,Lopata2009-1,Lopata2009-2} Nevertheless, for a number of reasons, such an approach becomes difficult for modeling a network of molecules. 
The first problem  one encounters when trying to quarantine self-interaction is an increasing memory cost,  as needed to store the incident and scattered fields for each molecule. 
To reduce this memory cost,  a field-partitioning technique\cite{Deinega2014} was recently developed, whereby one divides the computational volume of the EM field into the total field  and the scattered field (TF-SF). 
If each quantum emitter is allowed only one optical transition pathway, (e.g., a TLS),  this technique can drastically reduce the memory cost of  a simulation. However, 
the second problem one encounters is the possibility of one atomic or molecular site hosting more than two electronic states. In practice,
if each quantum emitter is a multi-level system and more than one optical transition pathway is allowed, it can become very difficult to model one specific optical transition pathway: now one must avoid the self-interaction from this pathway but include the self-interaction from other pathways coming from the same quantum emitter (which is effectively a one-site multiple scattering event\cite{Meystre2007}). Thus, one must distinguish between the scattered fields arising from different pathways  at the same spatial position (where the emitter lies), which violates the entire premise of the field-partitioning technique (which requires that only one type of field is defined at one point in space).
Thus, implementing an efficient Maxwell--Bloch scheme is difficult with field partitioning.
Alternatively, another option is the symmetry-adapted averaging technique\cite{Deinega2014} for canceling  self-interaction; however, this technique is computationally less stable than the field-partitioning technique. 
Recently, a photon Green functions (GFs) formulation was proposed, according to which  self-interaction can be excluded by carefully evaluating the real-part of GFs; however, this method is currently limited to weak excitations.\cite{Schelew2017}

To sum up, the advantage of the Maxwell--Bloch approach is that the added dissipative term is linear, so that a linear and stable Liouville equation can be simulated. The disadvantage of the Maxwell--Bloch approach is that, in practice, one needs to exclude the self-interaction from the EM field operating on each molecule, and when a network of molecules is considered (especially for multi-level systems), exclusion of self-interaction is nontrivial.

\subsection{Ehrenfest+R dynamics}
When encountering these difficulties as far as excluding self-interaction while keeping  a linear dissipative term, one is tempted to change strategy: why not first evaluate the nonlinear dissipative effect of self-interaction, and then second add another nonlinear dissipative term to the  Liouville equation to mimic solely the effect of vacuum fluctuations?
This approach should also allow one to recover the correct uniform FGR rate of spontaneous emission and is the philosophy behind Ehrenfest+R dynamics.

Let us now be more precise mathematically.  For spontaneous emission with a TLS (state $\ket{1}$ and $\ket{2}$), the semiclassical self-interaction in Ehrenfest dynamics leads to a decay rate ($\kEh$) proportional to the ground-state population\cite{Stroud1970,Li2018Spontaneous,Chen2018Spontaneous} ($\rho_{11}$):
\begin{equation}\label{eq:kEh-kFGR}
\kEh = \rho_{11} \kFGR
\end{equation}
where $\kFGR$ denotes the FGR decay rate. 
By virtue of Eq. (\ref{eq:kEh-kFGR}), we know that the rate of decay as arising from vacuum fluctuations ($\kR^{\text{vac}}$) must be of the form 
\begin{equation}\label{eq:kR-kFGR}
\kR^{\text{vac}} = \left(1 - \rho_{11}\right) \kFGR
\end{equation}
According to Ehrenfest+R, we will add another dissipative term (analogous to Eq. (\ref{eq:kR-kFGR})) to the Liouville equation while also taking care of energy conservation.  Already, we have shown that such an Ehrenfest+R approach correctly captures spontaneous emission\cite{Chen2018Spontaneous} and Raman scattering\cite{Chen2018Raman}. Moreover, by enforcing energy conservation, Ehrenfest+R can quantitatively distinguish coherent and incoherent scattering as produced during spontaneous emission from an arbitrary initial state\cite{Chen2018Spontaneous}. Furthermore, compared with previous approaches for excluding self-interaction, Ehrenfest+R should be easy to apply to a network of multi-level molecules with minimum memory cost for the field variables; after all, only one total E-field and B-field are necessary. Thus, in the near future, one of our goals is to use Ehrenfest+R to study a model of electrodynamics with multiple sites and multiple electronic states per site. Nevertheless, for the moment, among the benchmarking tests of Ehrenfest+R above, there is one clear omission. Namely, using FGR to model spontaneous emission assumes spontaneous emission is decoupled from all other dynamical processes. Thus, it is unknown whether Ehrenfest+R can quantitatively recover coherent and incoherent scattering in the limit of reasonably strong cw fields where electronic populations are oscillating rapidly on the time scale of spontaneous emission.  Our first goal in this paper is to provide benchmarks for answering this question: how will Ehrenfest+R perform for a TLS subject to not weak EM fields.\cite{footnote2}

In the context of this question,  we expect that the key issue of energy conservation must arise within classical and semiclassical approaches. Standard classical EM theory as well as the coupled Maxwell--Bloch equations do not satisfy energy conservation; and though these ansatzes make sense in the  linear regime with low incoming intensities and small excited state populations, one must wonder  if/how a lack of energy conservation shows its face when modeling optical signals beyond the linear regime. Thus, our second and more general goal for  this paper is to compare the performance of quantum, classical and especially semiclassical methods for modeling stimulated emission, paying special attention to energy conservation (which is not standard in most EM treatments).

\subsection{Outline and notation}
This paper is organized as follows. In Sec. \ref{sec:model}, we introduce the simple model TLS  we will  study. In Sec. \ref{sec:methods}, we introduce five methods for simulating light--matter interactions. In Sec. \ref{sec:numerical_details}, we give all the simulation details; see also the Appendix. In Sec. \ref{sec:results}, we carefully examine our simulation results, and compare and contrast different approaches. In Sec. \ref{sec:dicussion}, we discuss the accuracy of energy conservation and highlight why enforcing energy conservation is crucial for all semiclassical algorithms. We conclude in Sec. \ref{sec:conclusion}.

For notation, we use the following conventions: $\hbar\omega_0$ represents the energy difference between the excited state $\ket{e}$ and ground state $\ket{g}$, $\mu_{12}$ denotes the electric transition dipole moment, $\sigma$ denotes the width of molecule, $\Us$, $\UEM$ and $\Ut$ represent the energy of the quantum subsystem, the EM field and the total system, respectively. We work below in SI units, but will present all simulation results in dimensionless quantities  to facilitate conceptual understanding.

\section{Model}\label{sec:model}
To compare different methods clearly, we are interested in the following  model problem: an electronic two-level system (TLS) is coupled to an incident cw EM field in one dimension (say, along the $x$-axis)\cite{Allen1975}. Without loss of generality, we suppose the TLS is fixed at the origin. We further suppose the incident E-field is directed along the $z$-axis: 
\begin{equation}\label{eq:Ein}
\vEin(x, t) = E_0 \cos\left(kx - \omega t\right) \he_z
\end{equation}
Here, $E_0$ and $\omega$ are the amplitude and frequency of the cw EM field,  $\he_z$ is the unit vector along $z$-axis and $k = \omega /c$, where $c$ denotes the speed of light in vacuum.

The Hamiltonian for the TLS reads
\begin{equation}
\hH_s = \begin{pmatrix}
0 & 0 \\
0 & \hbar\omega_0
\end{pmatrix}
\end{equation}
in the basis of ground state $\ket{g}$ and excited state $\ket{e}$. $\hbar\omega_0$ is the energy difference between two states. We further suppose that the TLS has no permanent electric dipole moment and is coupled to the  E-field by the transition dipole moment. 
In general, the coupling between the incident E-field and the TLS can be written as  
\begin{equation}\label{eq:coupling-2}
\hat{V} = -\int d\vR \hP(\vR) \cdot \vEin(\vR, t)
\end{equation}
Here, $\vEin$ is expressed in Eq. (\ref{eq:Ein}) and the polarization density operator $\hP(\vR)$ is defined as
\begin{equation}\label{eq:P}
\hP(\vR) = \vxi(\vR)
\begin{pmatrix}
0 & 1 \\1 & 0
\end{pmatrix}
\end{equation}
where
\begin{equation}\label{eq:vxi}
\vxi(\vR) = \psi^{\ast}_{g}q\vR\psi_e =  \frac{\mu_{12}}{\sqrt{2\pi} \sigma} \exp(-x^2/2\sigma^2) \he_z
\end{equation}
denotes the polarization density of the TLS in 1D. In Eq. (\ref{eq:vxi}), we assume $\ket{g}$ is an $s$-orbital, $\ket{e}$   is a $p_z$ orbital,   $\sigma$ denotes the width of the electronic wave functions,  $\mu_{12} = |\bra{g}q\vR\ket{e}|$ is the magnitude of the transition dipole moment, and $q$ is the effective charge of the TLS. For convenience, we assume that the orientation of the transient dipole is parallel with the incident E-field (both along the $z$-axis). 

In realistic calculations, one frequently makes the point-dipole approximation, assuming that the length scale of the TLS is much less than the wavelength of the incident wave, i.e., $\hP(\vR) \rightarrow \hmu \vdelta(\vR)$, where we define $\vdelta(\vR) = \delta(x)\he_z$. In terms of the
transition dipole operator $\hmu$,
the coupling in Eq. (\ref{eq:coupling-2}) can be written as 
\begin{equation}\label{eq:coupling-1}
\hat{V} = -\hmu \cdot \vEin(\mathbf{0}, t)
\end{equation}
where 
\begin{equation}\label{eq:hmu}
\hmu = \mu_{12} \he_z
\begin{pmatrix} 
0 & 1 \\1 & 0
\end{pmatrix} 
\end{equation}
In order to quantify the magnitude of this coupling, a dimensionless quantity $\Omega/\kFGR$ is frequently used\cite{Mollow1969, Meystre2007}, where 
\begin{equation}\label{eq:rabi_frequency}
\Omega = \frac{\mu_{12} E_0}{\hbar}
\end{equation}
is called the Rabi frequency\cite{Knight1980} and 
\begin{equation}\label{eq:kFGR}
\kFGR = \frac{\omega_0}{\hbar\epsilon_0 c} |\mu_{12}|^2
\end{equation}
is the FGR decay rate  for the TLS in one dimension\cite{Astafiev2010, Li2018Spontaneous}. $\Omega/\kFGR \ll 1$ represents the weak coupling regime and $\Omega/\kFGR \gg 1$ represents the strong coupling regime.

Throughout this work, we will make the point-dipole approximation and we limit ourselves to discussion of a single TLS, so that using $\hP(\vR)$ or $\hmu$ will not change the results. However, for historical reasons (i.e., so that we may be compatible with most references), below we will use $\hmu$ when discussing the  coupled Maxwell--Bloch equations and the optical Bloch equation (OBE), while we will use $\hP(\vR)$ when discussing Ehrenfest dynamics and the Ehrenfest+R approach. Note that if we use the more general notation $\hP(\vR)$ (instead of $\hmu$), one can generalize the problem of light--matter interactions from one TLS to multiple TLSs at different positions $\vR$ without changing the form of the equations of motion.

\section{Methods}\label{sec:methods}

As discussed in the introduction, many methods have been proposed to model light--matter interactions. Here, we are interested in the following five methods, each of which treats self-interaction and vacuum fluctuations  differently: (i) the  coupled Maxwell--Bloch equations, (ii) the OBE,  (iii) Ehrenfest dynamics, (iv) the Ehrenfest+R approach and (v) classical dielectric theory (CDT). Apart from the newly developed Ehrenfest+R approach, all other methods are widely applied in different areas of chemistry, physics, and engineering. For instance, the OBE is widely applied in quantum optics and quantum information, the coupled Maxwell--Bloch equations and Ehrenfest dynamics are used to simulate laser experiments, and CDT is routinely applied in engineering and optics.

\subsection{Coupled Maxwell--Bloch equations: a double-counting of self-interaction}
One of the most widely applied methods to model light--matter interactions is the coupled Maxwell--Bloch equations\cite{Castin1995,Ziolkowski1995}. For the dynamics, the matter side is described  by the density operator $\hat{\rho}$, which is propagated quantum-mechanically:
\begin{equation}\label{eq:maxwell_bloch_rho}
\frac{d \hat{\rho} }{dt}  = - \frac{i}{\hbar} \left[ \hH_s - \hmu \cdot \vE(\mathbf{0}, t), \hat{\rho}  \right] + \Lrho{SE} 
\end{equation}
 Here, the phenomenological dissipative term $\Lrho{SE}$ reads
\begin{equation}\label{eq:Lrho_optical_bloch}
\Lrho{SE} \equiv \kFGR
\begin{pmatrix}
\rho_{22} & -\frac{1}{2}\rho_{12} \\
-\frac{ 1 }{2}\rho_{21} & -\rho_{22}
\end{pmatrix}
\end{equation}
$\Lrho{SE}$ describes the overall effects of the quantum field (self-interaction + vacuum fluctuations), which can be derived from quantum calculations, i.e., from a Lindblad term\cite{Wolf2008,Meystre2007} for an open quantum system. Note that in Eq. (\ref{eq:Lrho_optical_bloch}), the diagonal decay rate is called the \textit{population relaxation rate} ($\kFGR$) and the off-diagonal decay rate is called the \textit{dipole dephasing rate} ($\kFGR/2$). For spontaneous emission in a secular approximation, the dipole dephasing rate is half of population relaxation rate. However, for realistic systems, these two rates do not necessarily satisfy this relation and can be adjusted empirically.\cite{Boyd2008}

As far as  the EM field, all dynamics obey Maxwell's equations. The E-field is composed of the incident field plus scattered field generated by the TLS itself,
\begin{equation}\label{eq:E_composition}
\vE = \vEin + \vEscatt
\end{equation}
From Eq. (\ref{eq:E_composition}), to propagate $\vE$, since the explicit form of $\vEin$ at different times is given in Eq. (\ref{eq:Ein}), we need only to propagate $\vEscatt$:
\begin{subequations}\label{eq:maxwell_bloch_EM}
	\begin{align}
	\frac{\partial}{\partial t}\vB_{\text{scatt}}(\vR, t) &= -\vnabla \times \vEscatt(\vR, t) \\
	\frac{\partial}{\partial t}\vEscatt(\vR, t) &= c^2 \vnabla \times \vB_{\text{scatt}}(\vR, t) - \frac{\vJ(\vR, t)}{\epsilon_0}  \label{eq:E_field}
	\end{align}
\end{subequations} 
where $\epsilon_0$ denotes the vacuum permittivity and the current density $\vJ$ is calculated by the mean-field approximation
\begin{equation}\label{eq:J}
\vJ(\vR, t) = \frac{d \vPtot}{dt}  \vdelta(\vR) = \frac{d }{dt}\tr{\hat{\rho}\hmu}  \vdelta(\vR)
\end{equation}
Note that, according to Maxwell--Bloch, the E-field influences the electronic dynamics through the commutator $\left[\hH_s - \hmu \cdot \vE(\mathbf{0}, t), \hat{\rho}  \right]$, where the E-field is expressed in Eq. (\ref{eq:E_composition}).
Because this commutator obviously includes the effect of self-interaction, and yet the spontaneous emission rate $\kFGR$ accounts for both self-interaction and vacuum fluctuations in the $\Lrho{SE}$ term on the right hand side (RHS) of Eq. (\ref{eq:maxwell_bloch_rho}), Maxwell--Bloch evidently double-counts self-interaction.

\subsubsection{Advantages and disadvantages}
Because the scattered field is explicitly propagated, the advantage of the coupled Maxwell--Bloch equations is that one can model not only a single site, but also many quantum emitters as found in the condensed phase. That being said, however, this method double-counts self-interaction, leading to nonphysical results in both the electronic decay rate and the optical signals, which will be shown in this paper. More generally, stable and fast techniques are needed to separate self-interacting fields from otherwise incident fields in order to avoid double-counting. 
 
\subsection{The Classical Optical Bloch equation: exclusion of self-interaction in the EM-field}\label{sec:optical_bloch}
To exclude the self-interaction in Eq. (\ref{eq:maxwell_bloch_rho}), one needs to replace $\vE$ by $\vEin$ in the commutator on the RHS of  Eq. (\ref{eq:maxwell_bloch_rho}), resulting in the following Liouville equation
\begin{equation}\label{eq:optical_bloch}
\frac{d \hat{\rho} }{dt}= - \frac{i}{\hbar} \left[ \hH_s - \hmu \cdot \vEin(\mathbf{0}, t), \hat{\rho}  \right] + \Lrho{SE}
\end{equation} 
For the present paper,
$\vEin$ is defined in Eq. (\ref{eq:Ein}); one propagates Eqs. (\ref{eq:maxwell_bloch_EM}) and (\ref{eq:optical_bloch}) to obtain the dynamics of $\vEscatt$. When multiple sites are considered, one would need to distinguish between the incident and scattered fields for each site, which increases the complexity of the EM propagation scheme dramatically, just as for the coupled Maxwell--Bloch equations. However, for a single TLS, Eqs. (\ref{eq:maxwell_bloch_EM}-\ref{eq:optical_bloch}) form an efficient approximation known as the classical OBE\cite{Cohen-Tannoudji1998}. 


\subsubsection{Advantages and disadvantages}
The advantage of the OBE is its accuracy and solvability. This technique can provide useful analytical results, including, for example, the steady state solution of $\hat{\rho}$ and the susceptibility of molecule when exposed to a cw field.  As mentioned above, the disadvantage of the OBE is the implementational difficulty distinguishing the incident and scattered fields for each site when a large system (with multiple sites) are considered; this inefficiency is exactly the same problem as for the coupled Maxwell--Bloch equations.

Before concluding this subsection, we must re-emphasize the obvious: Eqs. (\ref{eq:maxwell_bloch_EM}a-b) are  the \textit{classical} equations of motion (i.e. Maxwell's equations) for a classical EM field, which is why Eqs. (\ref{eq:maxwell_bloch_EM}) and (\ref{eq:optical_bloch}) constitute the \textit{classical} OBE.  Within the quantum optics community, when field strength is large, one usually does not consider a classical EM field, so that one never propagates Eqs. (\ref{eq:maxwell_bloch_EM}a-b), and instead uses the \textit{quantum} OBE. According to such the \textit{quantum} OBE, 
one first propagates the matter quantum-mechanically with Eq. (\ref{eq:optical_bloch}), and second one calculates the intensity of the E-field at point $\vR$ quantum-mechanically by evaluating the correlation function for the matter degree of freedom\cite{Cohen-Tannoudji1998}. For example, for the TLS in our model, the intensity at point $\vR$ at time $t$ becomes
\begin{equation}\label{eq:E2_quantum}
\avg{I(t)} = \avg{\hE^{(-)}(\vR, t)\hE^{(+)}(\vR, t)} \propto \avg{\hat{S}_{+}(t- \frac{\vR}{c})\hat{S}_{-}(t- \frac{\vR}{c})}
\end{equation}		
where $\hE^{(+)}$ and $\hE^{(-)}$ represent the positive and negative frequency components of operator $\hE \equiv \hE^{(+)} e^{i\omega t} + \hE^{(-)}e^{-i\omega t}$, $\hat{S}_{+} \equiv e^{-i\omega t}\ket{e}\bra{g}$ and $\hat{S}_{-} \equiv e^{i\omega t}\ket{g}\bra{e}$. Similar expressions can be found for the B-field.  

For this paper,  we will mostly restrict ourselves to the classical rather than quantum OBE; we wish to evaluate comparable classical and semiclassical approaches without quantized photons. Nevertheless, in Fig. \ref{fig:2} below, we will compare Ehrenfest+R dynamics to the quantum OBE in the discussion section, when we investigate the ratio of coherent to incoherent EM intensity (and it would not be fruitful to consider the classical OBE). In general, the quantum OBE approach operates today as the standard  treatment for describing the dynamics of a TLS coupled to the radiation field.\cite{Zheng2003,Toyli2016}

\subsection{Ehrenfest dynamics: including self-interaction and ignoring vacuum fluctuations}\label{sec:Methods_Ehrenfest}

Ehrenfest dynamics are a semiclassical approach to electrodynamics derived from the full  Power-Zienau-Woolley quantum Hamiltonian after invoking the Ehrenfest (mean-field) approximation for both matter and photons\cite{May2011,Mukamel1999}. 
According to Ehrenfest dynamics, the semiclassical Hamiltonian reads
\begin{equation}\label{eq:Ehrenfest_Hsc}
\hH_{sc} = \hH_s - \int d\vR\  \hP(\vR) \cdot \vE_{\perp}(\vR, t)
\end{equation}
and the full dynamics are defined by
\begin{subequations}\label{eq:Ehrenfest}
	\begin{align}
	\frac{d \hat{\rho} }{dt} & = - \frac{i}{\hbar} \left[ \hH_{sc}, \hat{\rho}  \right] \label{eq:Ehrenfest_rho}\\
	\frac{\partial}{\partial t}\vB(\vR, t) &= -\vnabla \times \vE_{\perp}(\vR, t) \\
	\frac{\partial}{\partial t}\vE_{\perp}(\vR, t) &= c^2 \vnabla \times \vB(\vR, t) - \frac{\vJ_{\perp}(\vR, t)}{\epsilon_0}  \label{eq:Ehrenfest_E_field}
	\end{align}
\end{subequations} 
Here, $\hP$ is the polarization density operator; see Eq. (\ref{eq:P}). As mentioned above, after invoking the point-dipole approximation (i.e., $\hP(\vR) = \hmu \vdelta(\vR)$),  $\int d\vR \ \hP(\vR) \cdot \vE(\vR, t) = \hmu \cdot \vE(\mathbf{0}, t)$, so that Eq. (\ref{eq:Ehrenfest_Hsc}) is equivalent to the form of coupling in Eq. (\ref{eq:maxwell_bloch_rho}). $\vE_{\perp}$ ($\vJ_{\perp}$) denotes the transverse E-field (current density). For a single site, one can usually just neglect the $\perp$ nuance in Eqs. (\ref{eq:Ehrenfest_Hsc}-\ref{eq:Ehrenfest}).

Note that due to the lack of explicit dissipation, we can propagate Ehrenfest's  electronic dynamics with  a wave function formalism instead of with a density operator. In other words, we can replace Eq. (\ref{eq:Ehrenfest_rho}) by
\begin{equation}\label{eq:Ehrenfest_C}
\frac{d\vC}{dt} = -\frac{i}{\hbar} \hH_{sc} \vC
\end{equation}
For a TLS, $\vC = \left(c_1, c_2\right)$, where $c_1$ ($c_2$) is the quantum amplitude for the ground state (excited state). In realistic simulations, it is always more computationally efficient to propagate $\vC$ rather than $\hat{\rho}$.

From Eq. (\ref{eq:Ehrenfest_rho}), in capturing the quantum nature of radiation field, Ehrenfest dynamics consider only the self-interaction induced by the scattered field; one neglects the effect of vacuum fluctuations on the Liouville equation (i.e., there is no explicit dissipative term), which causes problems when describing spontaneous emission. In other words, if $\rho = \bigl( \begin{smallmatrix}0 & 0\\ 0 & 1\end{smallmatrix}\bigr)$ and $\vE(\vR) = \vB(\vR) = \mathbf{0}$ at time zero, the electronic system will not relax according to Eqs. (\ref{eq:Ehrenfest}). Let us now investigate spontaneous emission in more detail.

\subsubsection{The analytical form of dissipation induced by self-interaction}\label{sec:Eh-dissipation}
To begin our discussion, we rewrite Eq. (\ref{eq:Ehrenfest_rho}) as
\begin{equation}\label{eq:Ehrenfest-Lform}
\frac{d}{dt}\hat{\rho} = \Lrho{s}  + \Lrho{E\textsubscript{in}} +  \Lrho{E\textsubscript{scatt}}
\end{equation}
where we denote
\begin{subequations}\label{eq:LsLEin}
	\begin{align}
	\Lrho{s} &\equiv - \frac{i}{\hbar} \left[ \hH_s, \hat{\rho}  \right]    \\
    \Lrho{E\textsubscript{in}} 	&\equiv \ \ 
	\frac{i}{\hbar} \left[\hmu \cdot \vEin(\mathbf{0}, t), \hat{\rho}  \right]  \\
	\Lrho{E\textsubscript{scatt}} 	&\equiv \ \ 
	\frac{i}{\hbar} \left[\hmu \cdot \vEscatt(\mathbf{0}, t), \hat{\rho}  \right] 
	\end{align}
\end{subequations}
Here, $\Lrho{s}$, $\Lrho{E\textsubscript{in}}$ and $\Lrho{E\textsubscript{scatt}}$ denote the evolution of $\hrho$ due to $\hH_s$, the incident field and the scattered field, respectively. While $\Lrho{s}$ and $\Lrho{E\textsubscript{in}}$ do not cause electronic relaxation explicitly, in Ehrenfest dynamics,
one can prove that $\Lrho{E\textsubscript{scatt}}$  is effectively a dissipative term that is similar to Eq. (\ref{eq:Lrho_optical_bloch}),
\begin{equation}\label{eq:Leh}
\Lrho{E\textsubscript{scatt}} = 
\begin{pmatrix}
\kEh\rho_{22} & -\gEh \rho_{12} \\
-\gEh\rho_{21} & -\kEh\rho_{22}
\end{pmatrix}
\end{equation}
See Appendix \ref{sec:Ehrho} for a  detailed derivation.
Over a coarse-grained time scale ($\tau$) satisfying $1/\omega_0 \ll \tau \ll 1/\kFGR$, the nonlinear population relaxation rate reads
\begin{eqnarray}\label{eq:kEh-coarse}
\overline{\kEh}(t) = \kFGR \frac{|\rho_{12}|^2}{\rho_{22}}
\end{eqnarray}
Since no non-Hamiltonian term appears in Ehrenfest dynamics (see Eq. (\ref{eq:Ehrenfest_rho})), purity is strictly preserved for each trajectory, i.e., $|\rho_{12}|^2 = \rho_{11}\rho_{22}$, and thus Eq. (\ref{eq:kEh-coarse}) is equivalent to the expression in Eq. (\ref{eq:kEh-kFGR}). Similarly, within the same coarse-grained average, the effective dipole dephasing rate reads
\begin{equation}\label{eq:gEh-coarse}
\overline{\gEh}(t) = \frac{\kFGR}{2} \left(\rho_{11} - \rho_{22}\right)
\end{equation}
When a TLS is weakly excited ($\rho_{22}\rightarrow0$), according to Eqs. (\ref{eq:kEh-coarse}) and (\ref{eq:gEh-coarse}), $\overline{\kEh}\rightarrow \kFGR$ and $\overline{\gEh} \rightarrow \kFGR/2$, and thus $\Lrho{E\textsubscript{scatt}}$ defined in Eq. (\ref{eq:Leh}) agrees with $\Lrho{SE}$ defined in Eq. (\ref{eq:Lrho_optical_bloch}). In other words, Ehrenfest dynamics describe almost exactly the same dynamics as the OBE in the weak excitation limit.\cite{footnote1}


\subsubsection{Advantages and disadvantages}
Ehrenfest dynamics explicitly propagate the total EM field and are obviously equivalent to the coupled Maxwell--Bloch equations without the "hard" dissipative term ($\Lrho{SE}$ in Eq. (\ref{eq:Lrho_optical_bloch})): both techniques are applicable to the condensed phase with many emitters. Near the ground state, Ehrenfest dynamics effectively predict the same results as the OBE for a single TLS (which the coupled Maxwell--Bloch equations do not achieve because of double-counting). Another advantage of Ehrenfest dynamics is the enforcement of energy conservation  (which the coupled Maxwell--Bloch equations do not satisfy); see Appendix \ref{sec:EnergyConserve}. The disadvantage of Ehrenfest dynamics is obvious: Ehrenfest cannot describe the dynamics correctly when the system is strongly excited ($\rho_{11} \rightarrow 0$), which is why one introduces the extra dissipation in Eq. (\ref{eq:maxwell_bloch_rho}) in the first place.

\subsection{The Ehrenfest+R approach: counting self-interaction and vacuum fluctuation separately}\label{sec:Methods_EhR}
We have recently proposed an \textit{ad hoc} Ehrenfest+R approach to improve Ehrenfest dynamics in the limit of large excitation out of the ground state (as applicable under strong EM fields).
With the  Ehrenfest+R approach, we  want not only to  describe the electronic dynamics correctly, but we  want also to describe the EM field correctly. The former is rather easy to implement: we  need simply to augment Ehrenfest dynamics by adding the difference between $\Lrho{SE}$ and $\Lrho{E\textsubscript{scatt}}$ to the Ehrenfest equation of motion for the quantum subsystem. The latter, however, is difficult to implement: quantum-mechanically, the EM fields are operators and are fundamentally different from c-numbers. For example, according to QED, $\avg{\hE^2} \geq \avg{\hE}^2$, but this difference cannot be recovered in any classical scheme if only one trajectory is simulated.
Now, if one wants to distinguish $\avg{\hE^2}$ and $\avg{\hE}^2$ in a semiclassical way,  the standard approach is to introduce a swarm of trajectories. By calculating $\avg{\hE^2}$ and $\avg{\hE}^2$
with an ensemble average over many trajectories, one can find different values, especially if there is phase cancellation.
Such quasi-classical techniques have long been used in semiclassical quantum dynamics\cite{Hall2014,Tully1990,Cotton2013,Cotton2016,Ben-Nun2000,Kim2008,Nassimi2010,Crespo-Otero2018a,Gu2017}. 
Within the context of coupled nuclear-electronic dynamics, all successful semiclassical approaches average dynamics over multiple trajectories (including, e.g., surface hopping\cite{Tully1990}, the symmetrical quasi-classical (SQC) method\cite{Cotton2013,Cotton2016}, multiple spawning\cite{Ben-Nun2000}, and the Poisson bracket mapping equation\cite{Kim2008,Nassimi2010}, etc; see Ref. \citenum{Crespo-Otero2018a} for a general review).

Let us now briefly review the operational procedure for Ehrenfest+R; a full description of this method can be found in Ref. \citenum{Chen2018Spontaneous}.  An overall flowchart of the algorithm for Ehrenfest+R is shown in Algorithm \ref{Algorithm:+R}:  for each trajectory, we assign a random phase $\phi^l \in \left[0, 2\pi\right)$, (which will be motivated later), we propagate Ehrenfest dynamics for a time step $dt$ (see Eqs. (\ref{eq:Ehrenfest})), and then we introduce  a nonlinear dissipative event --- the +R correction --- which forces $\hrho$ to decay with an overall FGR rate; since this correction leads to energy dissipation for the quantum subsystem, we also rescale the EM field at each time step $dt$ to conserve energy; finally we perform an ensemble average over trajectories to calculate $\avg{\hrho}$, $\avg{\vE}$ and $\avg{\vE^2}$. Explicit equations are provided in Appendix \ref{sec:EhR_procedure}.
\begin{algorithm}[H]
	\caption{Ehrenfest+R Algorithm}
	\label{Algorithm:+R}
	\begin{algorithmic}[1]
		\For {Traj. $l = 1 : N$}
		\State \texttt{Assign a random phase $\phi^l$}
		\For {$t = t_{begin} : dt : t_{end}$}
		\State \texttt{Propagate Ehrenfest dynamics \ \ [Eq. (\ref{eq:Ehrenfest})]}
		\State \texttt{+R correction for $\hrho$ (or $\vC$) to enforce FGR decay for both diagonal and off-diagonal elements of $\hrho$ \ \ [Eqs. (\ref{eq:rho_ehR}-\ref{eq:LR})]}
		\State \texttt{Rescale EM field to conserve energy \ \ [Eqs. (\ref{eq:Ehrenfest+R-rescaleEM1}-\ref{eq:alpha_beta})]}
		\EndFor
		\EndFor
		\State \texttt{Average over Trajectories for $\avg{\hrho}_l$, $\avg{\vE}_l$ and $\avg{\vE^2}_l$}
	\end{algorithmic}
\end{algorithm}

\subsubsection{Advantages and disadvantages} The advantages of Ehrenfest+R approach are obvious: (i) this method recovers the correct spontaneous decay rate, while the coupled Maxwell--Bloch equations and Ehrenfest dynamics cannot; (ii) by taking an average over a swarm of trajectories (with random values of $\phi^l$ determined for each trajectory at the start of the simulation), Ehrenfest+R not only conserves energy, but also distinguishes coherent scattering from incoherent scattering. The disadvantage of Ehrenfest+R is the computational cost necessitated by introducing a sampling of trajectories with different phases. However, in the benchmark work presented here, to obtain acceptable results, we find the Ehrenfest+R requires only on the order of $10^2$ trajectories. Our hope is that, for large systems, the cost of Ehrenfest+R will remain very moderate.

\subsection{Classical Dielectric Theory (CDT): a non-explicit double-counting of self-interaction}
Classical electrodynamics is always a competing approach for modeling light--matter interactions.
According to CDT, without any free charge, the displacement field $\vD$, the  auxiliary magnetic field $\vH$, the electric field $\vE$ and the magnetic  induction $\vB$ are all transverse, and 
the EM field obeys the classical Maxwell's equations:
\begin{subequations}\label{eq:Maxwell-classical-general}
	\begin{align}
	\frac{\partial}{\partial t}\vB(\vR, t) &= -\vnabla \times \vE(\vR, t)\\
	\frac{\partial}{\partial t}\vD(\vR, t) &= \frac{1}{\mu_0}\vnabla \times \vH(\vR, t)
	\end{align}
\end{subequations}
Here, as always, the equations that relate fields with and without matter are $\vD(\vR, t) = \epsilon_0 \vE(\vR, t) + \vPtot(\vR, t)$, $\vB(\vR, t) = \mu_0(\vH(\vR, t) + \mathbf{M}(\vR, t))$, where $\vPtot$ is the polarization field and $\mathbf{M}$ is the magnetisation field. If we assume a linear medium,
the constitutive relationships become:
\begin{subequations}\label{eq:Maxwell-classical-correspondance}
	\begin{align}
	\vH &= \frac{1}{\mu} \vB \\
	\vE &= \frac{1}{\epsilon} \vD \label{eq:E-classical-correspondance}
	\end{align}
\end{subequations}
Here, $\epsilon$ and $\mu$ denote the electric and magnetic permeabilities.   
Today, the most popular method for numerically solving Eqs. (\ref{eq:Maxwell-classical-general}-\ref{eq:Maxwell-classical-correspondance}) is the finite-difference time-domain (FDTD) method\cite{Taflove2005},
wherein the displacement field $\vD$ and the magnetize field $\vH$ are explicitly propagated in the time domain (instead of $\vE$ and $\vB$), and all vector fields are propagated with a Yee cell\cite{Yee1966}.

When one can ignore the magnetic interactions (as is true for our model with no magnetic susceptibility), one can propagate either $\vH$ or $\vB$. Since $\vH = \vB/\mu_0$, where $\mu_0$ denotes the vacuum magnetic permeability, Eqs. (\ref{eq:Maxwell-classical-general}-\ref{eq:Maxwell-classical-correspondance}) are reduced to
\begin{subequations}\label{eq:Maxwell-classical}
	\begin{align}
	\frac{\partial}{\partial t}\vH(\vR, t) &= -\frac{1}{\mu_0}\vnabla \times \vE(\vR, t)\\
		\frac{\partial}{\partial t}\vD(\vR, t) &= \vnabla \times \vH(\vR, t) \\
	\vD(\vR, \omega) &= \epsilon(\omega)\vE(\omega)
	\end{align}
\end{subequations}
Here,
$\vD$ and $\vE$ are connected by $\vD = \epsilon_0\vE + \vPtot$.
In general, the optical response of materials is described by the frequency-dependent dielectric function $\epsilon(\omega)$. By defining $\epsilon(\omega) = \epsilon_0(1 + \chi(\omega))$, one obtains
\begin{equation}\label{eq:PE-identity}
\vPtot(\omega) = \epsilon_0 \chi(\omega)\vE(\omega)
\end{equation}
As long as the dielectric function is given, in principle one can apply FDTD to propagate the $\vD$, $\vH$ and $\vPtot$ fields in the time domain (using Eqs. (\ref{eq:ddPddt_discrete}-\ref{eq:FDTD_Maxwell})).  For details see Sec. \ref{sec:numerical_details} and Appendix \ref{sec:fdtd}.

A derivation of the dielectric function for a TLS is well-known\cite{Mukamel1999,Boyd2008}, and the standard approach is rederived in Appendix \ref{sec:Chi}. Here, we present only the final expressions for $\chi(\omega)$. In the  weak-excitation limit (linear response regime), the dielectric function for a TLS is
\begin{equation}\label{eq:chi_L}
\chi^{\text{L}}(\omega) \approx \frac{\omega_p^2}{\omega_0^2 - \omega^2 - i\omega \kFGR}
\end{equation}
which corresponds to a Lorentz medium. Here, $\omega_p = \sqrt{2\mu_{12}^2\omega_0 /\epsilon_0\hbar}$. Beyond linear response, a nonlinear dielectric function can be expressed to the lowest nonlinear order in the series expansion of the incoming field (see Appendix \ref{sec:Chi}):
\begin{equation}\label{eq:chi_NL}
\chi^{\text{NL}}(\omega) \approx \chi^{\text{L}}(\omega) \left[1 - \frac{1}{1 + 4(\omega - \omega_0)^2/\kFGR^2 }  \frac{|E_0|^2}{|E_s|^2}  \right]
\end{equation}
In Eq. (\ref{eq:chi_NL}), $E_0$ is the amplitude of the incident wave defined in Eq. (\ref{eq:Ein}) and we define $|E_s|^2 \equiv \hbar^2 \kFGR^2 / 2 |\mu_{12}|^2$.
One interesting property of the series expansion leading to $\chi^{\text{NL}}$ is that the series converges only when $|E_0| / |E_s| \leq 1$. Therefore, this expansion cannot be used to model very strong light--matter interactions.

In Appendix \ref{sec:Chi}, we show CDT (Eq. (\ref{eq:Maxwell-classical})) with Eq. (\ref{eq:chi_L}) for the dielectric function double-counts self-interaction for a TLS because of a mismatch between the derivation of $\chi(\omega)$ (which assumes $\vPtot = \epsilon_0 \chi\vE_{\text{in}}$) and the way the polarization is used within CDT ($\vPtot = \epsilon_0 \chi\vE$, Eq. (\ref{eq:PE-identity})). In other words, CDT suffers the same problem effectively as the coupled Maxwell--Bloch equation. This double-counting becomes obvious if we evaluate the equation of motion for the optical polarization. See Table \ref{tbl:energy_conservation}.

\subsubsection{Advantages and disadvantages}
For CDT, one propagates only Maxwell's equations (i.e., no Schr\"odinger equation) in the time domain, and one can perform large-scale calculations within a parallel architecture. The disadvantage of CDT is that, by treating the matter side classically, one fails to capture any quantum features of the light--matter interactions, unlike the case for semiclassical simulations. Furthermore, CDT double-counts self-interaction for a TLS in a similar manner to the coupled Maxwell--Bloch equations.

\begin{table*}

	\caption{Synopsis of the main features for the five different approaches chosen for modeling light--matter interactions}
	\label{tbl:energy_conservation}
	\scriptsize
	\centering
	\begin{tabular}{lccr}
		\hline
		Approach & Recover SE& Equation of motion for optical polarization& Energy conservation\cite{footnote3}
		\\ \hline 
		Optical Bloch & True &  $\ddot{\vPtot}(t) + \kFGR \dot{\vPtot}(t) + \tilde{\omega}_0^2 \vPtot(t) = \epsilon_0\omega_p^2 W_{12}(t)  \vE_{\text{in}}(t)$ & True (QOBE) 
		\\
		\tiny
		Eqs. (\ref{eq:Lrho_optical_bloch}) and (\ref{eq:optical_bloch}) & & & / False (COBE)
		\\ \\
		Maxwell--Bloch & False &$\ddot{\vPtot}(t) + \kFGR \dot{\vPtot}(t) + \tilde{\omega}_0^2 \vPtot(t) = \epsilon_0 \omega_p^2 W_{12}(t) \vE(t)$& False \\ 
		\tiny
		Eqs. (\ref{eq:maxwell_bloch_rho}-\ref{eq:J}) & & &
		\\ \\
		Ehrenfest & False \tiny(True only &$\ddot{\vPtot}(t) + \omega_0^2 \vPtot(t) = \epsilon_0 \omega_p^2 W_{12}(t)\vE(t)$& True \\ 
		\tiny
		Eqs. (\ref{eq:Ehrenfest}) and (\ref{eq:J}) & when $\rho_{11}\rightarrow 1$) & &
		\\ \\
		Ehrenfest+R & True &$\ddot{\vPtot}(t) + 2\gR(t) \dot{\vPtot}(t) + \left[\omega_0^2 + \dot{\gamma}_{\text{R}}(t)   + \gamma_{\text{R}}^2(t)\right] \vPtot(t) = \epsilon_0 \omega_p^2 W_{12}(t)\vE(t)$ & True \\ 
		\tiny
		Appendix \ref{sec:EhR_procedure} & & &
		\\ \\
		CDT-Lorentz & --- &$\ddot{\vPtot}(t) + \kFGR \dot{\vPtot}(t) + \omega_0^2 \vPtot(t) = \epsilon_0 \omega_p^2  \vE(t)$ & False 
		\\
		\tiny
		Eqs. (\ref{eq:Maxwell-classical}-\ref{eq:chi_L}) & & & 
		\\ \hline
	\end{tabular}
	\normalsize

\end{table*}

\subsection{Summary of methods}
After introducing the five methods above, we now summarize the main features of each method in Table \ref{tbl:energy_conservation}, highlighting  (i) the ability to recover the FGR rate in spontaneous emission (SE); (ii) the effective equations of motion for the optical polarization ($\vPtot$), and (iii) whether or not energy is conserved. See Appendix \ref{sec:OpticalPolarization} for all derivations.

Note that in Table \ref{tbl:energy_conservation}, we define $\vPtot = \int d\vR \tr{\hrho\hP} = \tr{\hrho\hmu}$; we also define $\omega_p \equiv \sqrt{2\mu_{12}^2\omega_0 /\epsilon_0\hbar}$ as the plasmon frequency,  $W_{12} \equiv \rho_{11} - \rho_{22}$, $\vEin(t)$ and $\vE(t)$ are short for $\vEin(\mathbf{0}, t)$ and $\vE(\mathbf{0}, t)$. From the equations of motion of $\vPtot$ for each method, we can clearly ascertain whether a method double-counts self-interaction or not. For example, in the coupled Maxwell--Bloch equations as well as CDT, both a dissipative term $\kFGR\dot{\vPtot}$ and the total E-field appears, indicating a double-counting of self-interaction.

\section{Numerical Details}\label{sec:numerical_details}
Our parameters are chosen as follows (listed both in natural units $c = \hbar = \epsilon_0 = 1$ and $[t] = 1\times 10^{-17}$ s as well as in SI units): the energy difference of TLS is $\hbar\omega_0 = 0.25$ (16.5 eV), the transient dipole moment is $\mu_{12} = 0.025\sqrt{2}$ (11282 C$\cdot$nm/mol), the width of the TLS is $\sigma = 0.50$ (1.5 nm). We propagate Maxwell's equations on a 1D grid with  spatial spacing $\Delta x = 0.10$ (0.3 nm), time spacing $\Delta t = 0.05$ ($5\times 10^{-4}$ fs), and our spatial domain ranges from $x_{\text{min}} = -4\times 10^4$ (-12 $\mu$m) to $x_{\text{max}} = 4\times 10^4$ (12 $\mu$m). The propagation time is $t_{\text{max}} = 10^5$ (1 ps). We calculate the steady-state intensity of the EM field by  averaging the EM field generated in the time range $[t_{\text{max}} - t_0, t_{\text{max}}]$, where $t_0 = 10^4$ (100 fs). 

For the CDT simulation, we use FDTD with the standard Yee cell\cite{Yee1966}. For a Lorentz medium ($\chi^{\text{L}}(\omega)$), we propagate Eqs. (\ref{eq:Maxwell-classical}), (\ref{eq:PE-identity}) and (\ref{eq:ddPddt_fdtd}) simultaneously as is standard.\cite{Sullivan2013} For the nonlinear dielectric function $\chi^{\text{NL}}(\omega)$ in Eq. (\ref{eq:chi_NL}), since the incident field is monochromatic, we need simply to treat $\left[1 - \frac{1}{1 + 4(\omega - \omega_0)^2/\kFGR^2 }  \frac{|E_0|^2}{|E_s|^2}  \right]$ as a constant during the simulation so that the equations of motion are similar to the linear case. The standard trick for simulating dynamics with a Lorentz susceptibility is repeated in Appendix \ref{sec:fdtd}.

For all methods apart from CDT, all time derivatives for fields and matter are propagated by Runge-Kutta 4th-order solver\cite{Butcher2008} and spatial gradients are evaluated on a real space grid with a two-stencil. For the Ehrenfest+R approach, we average over 48 trajectories unless stated otherwise.

\section{Results} \label{sec:results}

In this section, we report the  electronic dynamics and the steady-state optical signals arising  when an incident cw field excites a TLS starting in the ground state.

\subsection{Electronic dynamics}
\begin{figure*}
	\includegraphics[width=\textwidth]{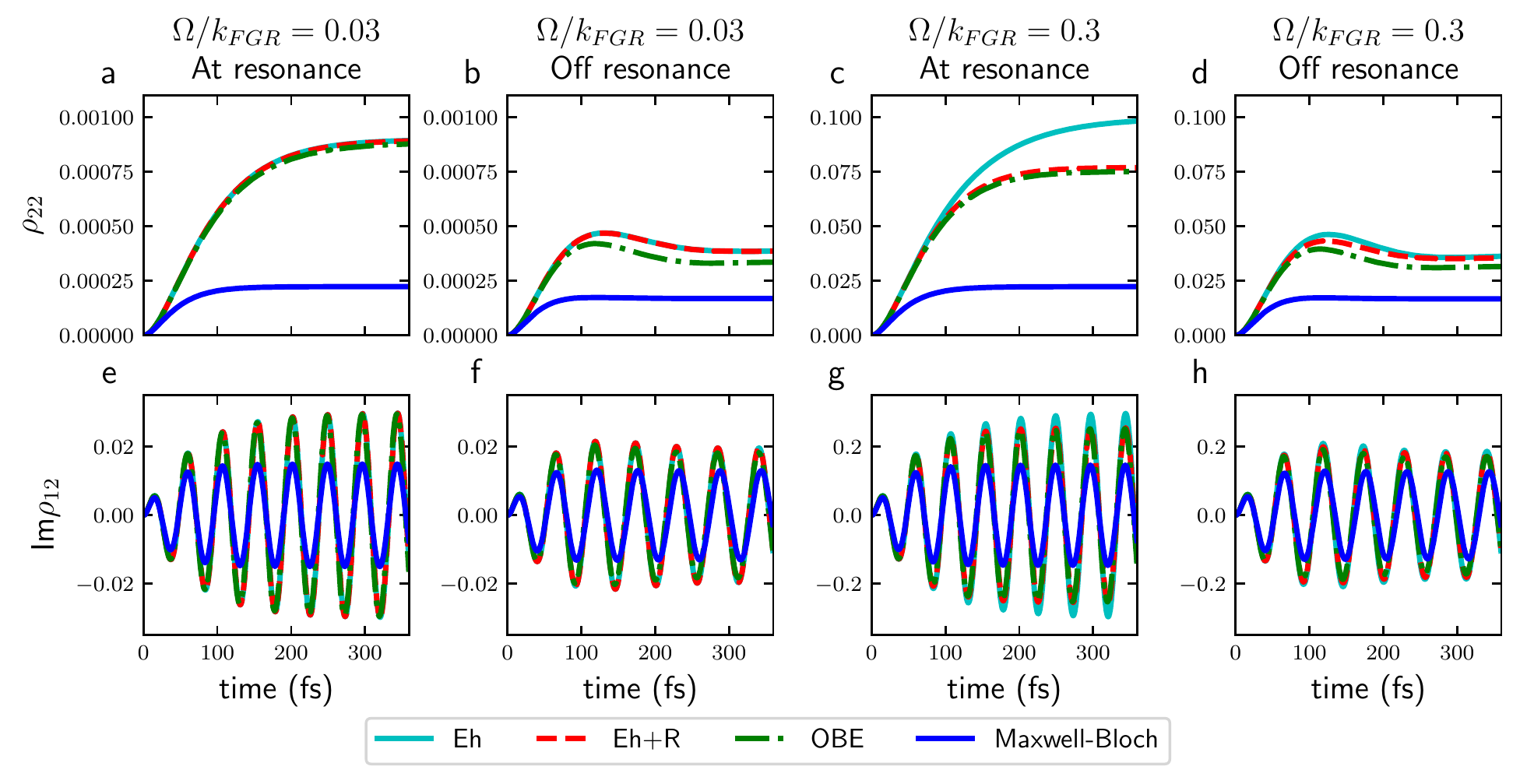}
	\caption{\label{fig:traj} Electronic dynamics of a TLS excited by an incident cw field: (upper) the excited state population ($\rho_{22}$) versus time, (bottom) the imaginary part of  coherence $\text{Im}\rho_{12}$ versus time.
	Four different conditions are plotted: (from left to right) weak at-resonant field ($\Omega/\kFGR = 0.03$, $\omega = \omega_0$), weak off-resonant field ($\Omega/\kFGR = 0.03$, $(\omega - \omega_0)/\kFGR=0.64$), slightly stronger at-resonant field  ($\Omega/\kFGR = 0.3$, $\omega = \omega_0$) and slightly stronger off-resonant field ($\Omega/\kFGR = 0.3$, $(\omega - \omega_0)/\kFGR=0.64$).
	In each subplot, four methods are compared: (i) Ehrenfest dynamics (cyan solid); (ii) the Ehrenfest+R approach (red solid); (iii) the OBE (dashed-dotted green) and (iv) the coupled Maxwell--Bloch equations (dashed blue). 
	Other parameters are listed in Sec. \ref{sec:numerical_details}. 
	Note that for the weak coupling case, Ehrenfest, Ehrenfest+R and OBE roughly agree. For the case of the stronger incident field, 
	Ehrenfest+R predicts almost the same results as the OBE; Ehrenfest dynamics overestimate the electronic response at resonance because the method ignores of vacuum fluctuations.  The coupled Maxwell--Bloch equations underestimate the electronic dynamics in all situations because of the double-counting of self-interaction.
	} 
\end{figure*}
Fig. \ref{fig:traj} shows the electronic dynamics of a TLS as a function of time for all methods (except CDT for which there are no explicit TLS dynamics). Among these methods, the OBE (green dashed-dotted) can be regarded as a "standard" method for all conditions. Our results are as follows.
\begin{enumerate}
	\item  For a weak at-resonant cw field ($\Omega/\kFGR = 0.03$, $\omega=\omega_0$), both Ehrenfest (cyan solid) and Ehrenfest+R (dashed red) quantitatively agree with the OBE for the evolution of the excited state population $\rho_{22}$ (Fig. \ref{fig:traj}a) and the imaginary part of $\text{Im}\rho_{12}$ (Fig. \ref{fig:traj}e).  Ehrenfest and Ehrenfest+R agree in the limit of weak excitation (where the +R correction [proportional to $\rho_{22}$] for Ehrenfest+R is negligible) and both methods agree with the correct OBE. That being said, the coupled Maxwell--Bloch equations (blue solid) predict different results: both $\rho_{22}$ and $\text{Im}\rho_{12}$ are drastically suppressed (see Figs. \ref{fig:traj}a,e) because of the double-counting of self-interaction.
	
	\item   For a weak off-resonant cw field ($\Omega/\kFGR = 0.03$, $(\omega-\omega_0)/\kFGR = 0.64$), not surprisingly, the dynamics for $\rho_{22}$ and $\text{Im}\rho_{12}$ are suppressed (Figs.  \ref{fig:traj}b,d, respectively) compared with the  resonant case. Interestingly,  both Ehrenfest and Ehrenfest+R predict a slightly higher response of $\rho_{22}$ compared with the OBE. This slight  difference originates from two factors. (i) The  effective Ehrenfest dissipative term $\Lrho{E\textsubscript{Scatt}}$ approaches $\Lrho{SE}$ near the ground state only in a coarse-grained sense. (ii) More importantly, the off-diagonal Ehrenfest term $\Lrho{E\textsubscript{Scatt}}$ is purely imaginary, which leads to slightly less dephasing compared with the $\Lrho{SE}$ term from the OBE, which has a real part, e.g., $\kFGR\rho_{12}/2$. Thus, the OBE and Ehrenfest do not yield the exact same dynamics even though the  absolute values of the off-diagonal components of both $\Lrho{E\textsubscript{Scatt}}$ and  $\Lrho{SE}$  are identical. See Appendix \ref{sec:Ehrho} for a detailed discussion. Again, Maxwell--Bloch disagrees with all of the other methods because of double-counting.
	
	\item When the cw field is amplified to slightly beyond the weak coupling limit ($\Omega/\kFGR = 0.3$), the new feature that arises is that Ehrenfest  now over-responds both for $\rho_{22}$ and $\rho_{12}$ compared with the OBE; at the same time, Ehrenfest+R still nearly agrees with the OBE, just as in the weak coupling case. Obviously, the inclusion of vacuum fluctuations becomes more and more important as the amplitude of the incident field  and  the excited state population increases. For strong fields, Ehrenfest+R becomes an important correction to Ehrenfest dynamics.
\end{enumerate}

\subsection{Steady-state optical signals}
\begin{figure*}
	\includegraphics[width=0.8\textwidth]{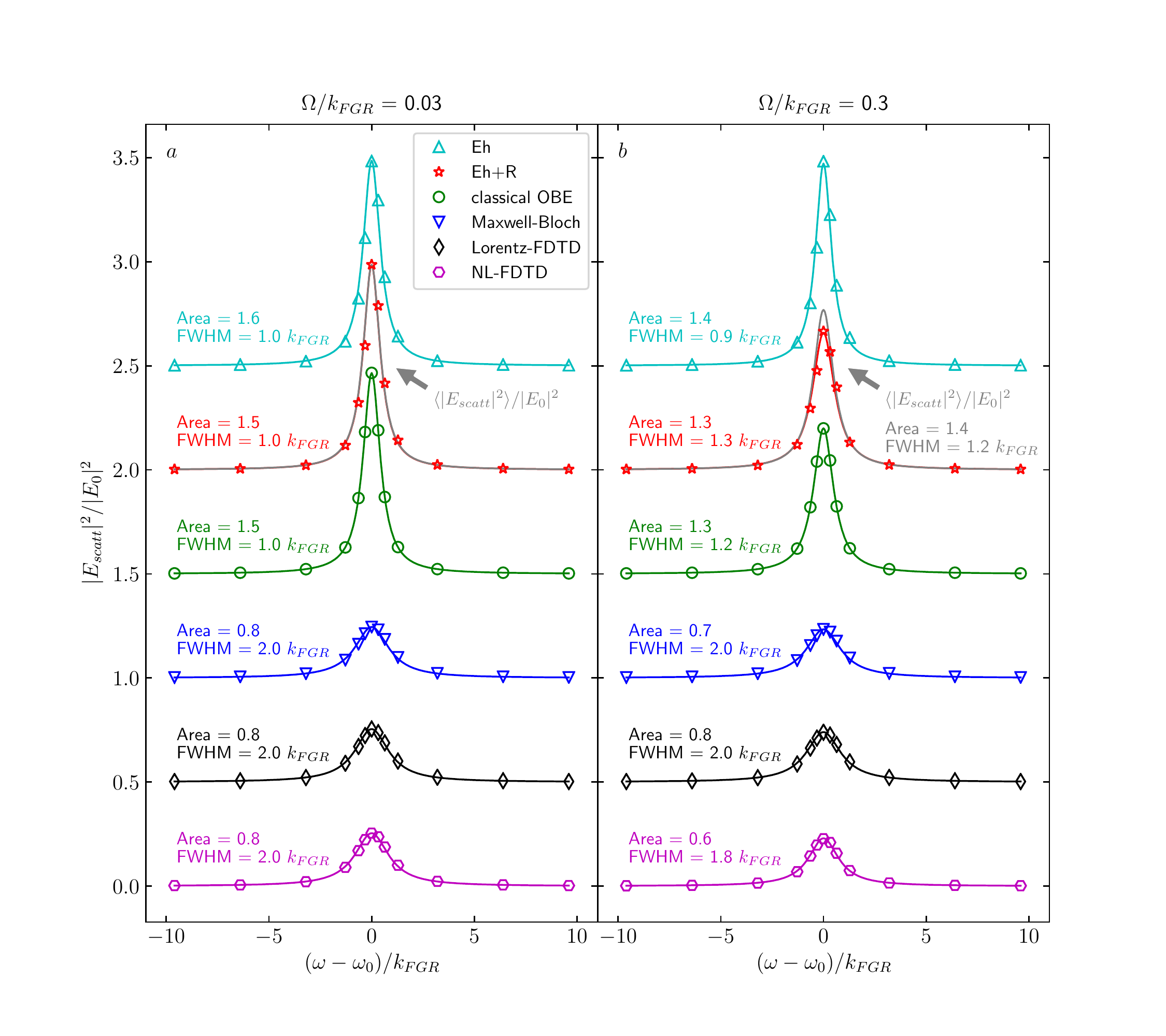}
	\caption{\label{fig:1} Steady-state intensities for the scattered E-field for a TLS as a function of frequency of the incident cw field: (left) weak incident field ($\Omega/\kFGR = 0.03$) and (right) slightly stronger incident field ($\Omega/\kFGR = 0.3$). Six methods are compared (from top to bottom): (i) Ehrenfest dynamics (cyan), (ii) the Ehrenfest+R approach (red for coherent scattering and grey for total scattering intensity), (iii) the  OBE (green), (iv) the coupled Maxwell--Bloch equations (blue), (v) CDT with a Lorentz medium (black, $\chi^{\text{L}}(\omega)$ given in Eq. (\ref{eq:chi_L}) ), and (vi) CDT with a non-linear medium (purple, $\chi^{\text{NL}}(\omega)$ given in Eq. (\ref{eq:chi_NL})). 
	Simulation data (dots) are fitted to a Lorentzian function defined in Eq. (\ref{eq:Lorentzian}), and the fitted parameters are also labeled (integral area and FWHM of Lorentzian). Simulation parameters are listed in Sec. \ref{sec:numerical_details}.
		Note that in the linear response regime (left), Ehrenfest, Ehrenfest+R agree with the OBE,  while  Maxwell--Bloch  and CDT predict different results because of the double-counting of self-interaction. Beyond linear response (right), Ehrenfest dynamics overestimate the intensity and underestimate the FWHM because of the absence of vacuum fluctuations; Ehrenfest+R and the OBE predict the correct trends: less intense and broader peaks, which are known as saturation effects; for Maxwell--Bloch or CDT, these tendencies are not obvious. Finally, when only including only the third-order nonlinear term, the performance of CDT is not enhanced, as the method still does not capture the broadening correctly.
	} 
\end{figure*}
In Fig. \ref{fig:1}, we plot the steady-state intensity of the scattered field ($|E_{\text{scatt}}|^2/ |E_0|^2$) as a function of incident cw wave frequency ($(\omega - \omega_0)/\kFGR$) when the light--matter coupling ($\Omega/\kFGR$) is weak ($\Omega/ \kFGR = 0.03$, left) and relatively strong ($\Omega/\kFGR = 0.3$, right). Note that we plot the overall scattered field, integrated over all intensities. 
While the dots in Fig. \ref{fig:1} represent the simulation data with specific incident frequencies $\omega$, we also  fit these data to a  Lorentzian in order to better capture the line width and the magnitude of the optical signal; the Lorentzian is defined as 
\begin{equation}\label{eq:Lorentzian}
f(\omega) = \frac{A}{\pi}\frac{\frac{1}{2}\Gamma}{(\omega - \omega_0)^2 + \left(\frac{1}{2}\Gamma\right)^2}
\end{equation}
where $A$ denotes the total integrated area of $f(\omega)$ and $\Gamma$ denotes full width at half maximum (FWHM) of $f(\omega)$.

For the weak coupling ($\Omega/ \kFGR = 0.03$; see Fig. \ref{fig:1}a), just as for electronic dynamics (see Fig. \ref{fig:traj}), Ehrenfest (cyan) and Ehrenfest+R (red) agree with the OBE (green) while both Maxwell--Bloch (blue)  and CDT (black for linear $\chi$ and purple for nonlinear $\chi$) predict different results. When the excitation is weak, both Ehrenfest, Ehrenfest+R and the OBE correctly predict the FGR rate for the electronic relaxation, which now becomes the FWHM for the lineshape. Due to the double-counting of self-interaction, however, the coupled Maxwell--Bloch equations and CDT predict twice the  correct FWHM; see the detailed discussion in Appendix \ref{sec:OpticalPolarization}.  Lastly, we note that, if one wants to use CDT to predict the correct FWHM for a TLS, one can reduce the width of the dielectric function in half (i.e., reduce $\kFGR$ to $\kFGR/2$ in Eq. (\ref{eq:chi_L})). Such a result can  indeed be reproduced if one takes care  to avoid double-counting. However, we emphasize that generalizing this result to large quantum subsystems (e.g., beyond a TLS) is either tedious or impossible. For this reason, in this paper, we have used the standard susceptibility for a TLS, i.e., Eqs. (\ref{eq:chi_L}-\ref{eq:chi_NL}) as found in Refs. \citenum{Mukamel1999} and \citenum{Boyd2008}.

Now let us move to a slightly stronger cw field ($\Omega/ \kFGR = 0.3$).  See Fig. \ref{fig:1}b.
We do not choose a very large field because the nonlinear FDTD simulation will become unstable when $\Omega/\kFGR \rightarrow 1$ due to the convergence issue of $\chi^{\text{NL}}(\omega)$; see Eq. (\ref{eq:chiNL_series}). For even a moderately strong field ($\Omega/\kFGR = 0.3$), the OBE predicts a saturation effect, for which the intensity of the scattered field is suppressed and the FWHM is broadened. Similar tendencies can also be found in Ehrenfest+R. 

Interestingly, for the coupled Maxwell--Bloch equations, a saturation effect is not obvious because of  the double-counting of self-interaction. Furthermore, because Ehrenfest does not include vacuum fluctuations, this method predicts the exactly incorrect trend: the FWHM decreases for large incident fields. Finally, regarding CDT, it is not surprising at all that the  linear Lorentz medium results do not change when the incident field strength is increased. More interestingly, even if we include the lowest order of non-linearity, CDT predicts the incorrect trend for the absorption FWHM (just like Ehrenfest). Apparently,  including only the lowest order of non-linearity is not enough for an accurate description of optical signals outside of linear response. 

Finally, before concluding, we note that  Ehrenfest+R also makes a prediction of the total scattering intensity ($\avg{|\hE|^2}$).
As such, Ehrenfest+R differs from all the other methods presented in Fig. \ref{fig:1}, which predict only the intensity of coherent scattering. The only other method which can make prediction of coherent versus incoherent EM dynamics is the quantum OBE (see Sec. \ref{sec:optical_bloch}), which is considered the gold standard for modeling a quantum field of photons interacting with a TLS. Although we do not plot the quantum OBE results in Fig. \ref{fig:1}, we will compare Ehrenfest+R with the quantum OBE below in the Discussion.

\begin{figure}
	\includegraphics[width=0.5\textwidth]{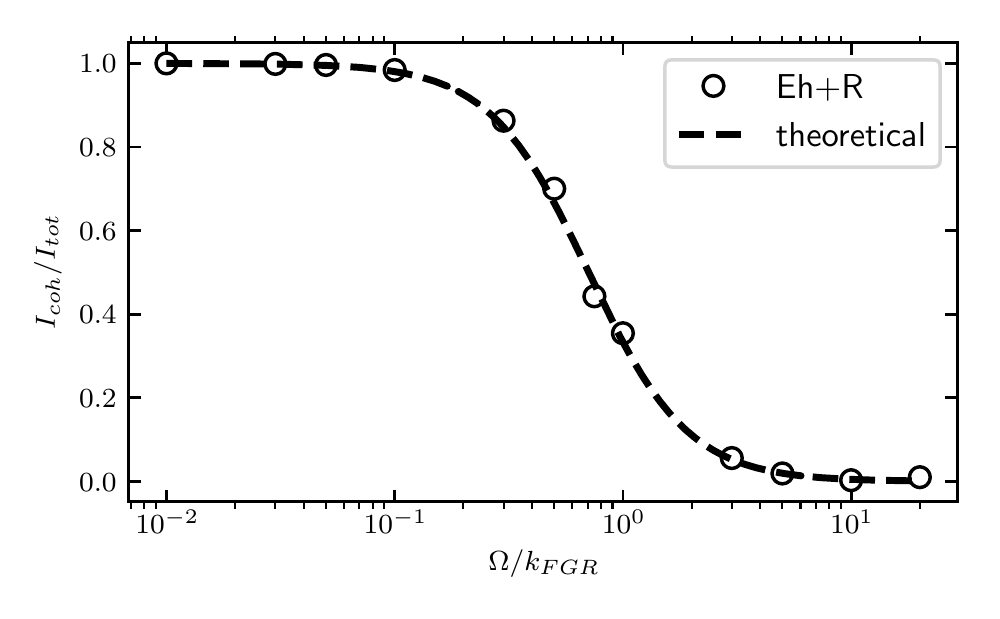}
	\caption{\label{fig:2} Ratio between the steady-state intensities of coherent and total scattered fields ($I_{\text{coh}} / I_{\text{tot}}$) as a function of light--matter coupling ($\Omega/\kFGR$) when a TLS is excited by an incident cw field at resonance ($\omega = \omega_0$). Ehrenfest+R result (open circle) is compared with the theoretical prediction (dashed line) in Eq. (\ref{eq:Icoh_over_Itot_Mollow}). 480 trajectories are averaged for Ehrenfest+R. All other parameters are the same as Fig. \ref{fig:1}. 
	} 
\end{figure}

\section{Discussion} \label{sec:dicussion}
The above results demonstrate that the Ehrenfest+R approach may be an advantageous method to model light--matter interactions: this method not only predicts similar electronic dynamics and the same coherent scattered field as  the classical OBE, but also directly models the total scattered field by enforcing energy conservation so that one can predict $\avg{\vE^2}$ and $\avg{\vE}^2$ independently. Thus, in the future, we believe we will be able to use the Ehrenfest+R approach to correctly model light--matter interactions for many emitters (without the requirement of excluding self-interaction, as needed for the coupled Maxwell--Bloch equations). Because so many collective optical phenomena have been recognized --- for example, resonant energy transfer\cite{Li2018Tradeoff,Power1983}\cite{Salam2018}, superradiance\cite{Gross1982,Spano1990,Lim2004} and quantum beats\cite{Thakkar2015} --- the Ehrenfest+R approach is a very tempting tool to generalize and apply.

That being said, it remains
to demonstrate that the Ehrenfest+R approach predicts the correct total scattered field as compared with the quantum OBE. To that end, in Fig. \ref{fig:2} we plot the ratio between the coherent and total scattered fields ($I_{\text{coh}} / I_{\text{tot}}$) for a wide range of field intensities ($\Omega/\kFGR$)  for the case that the TLS is excited at resonance  by an incident cw field ($\omega = \omega_0$). From Fig. \ref{fig:2}, results for Ehrenfest+R (open circle) quantitatively match  the theoretical prediction calculated by Mollow\cite{Mollow1969} (dashed line):
\begin{equation}\label{eq:Icoh_over_Itot_Mollow}
\frac{I_{\text{coh}}}{I_\text{tot}} = \frac{(\omega - \omega_0)^2 + \frac{1}{4}\kFGR^2}{\frac{1}{2}\Omega^2 + (\omega - \omega_0)^2 + \frac{1}{4}\kFGR^2}
\end{equation}
The quantitative agreement between Ehrenfest+R and the analytical calculation based on the quantum OBE in Fig. \ref{fig:2} can be understood as follows.
From the procedure of Ehrenfest+R (see Sec. \ref{sec:Methods_EhR}), we are guaranteed  both that the excited state will relax with the FGR rate and that energy will be conserved. On the one hand, because of the correct electronic relaxation, Ehrenfest+R  must predict nearly the same electronic dynamics as the OBE, so that the coherent scattered field predicted by Ehrenfest should be accurate. On the other hand, because energy is conserved at each time step, the total intensity of the scattered field has to be correct. By combining these two sides, it is not surprising that Ehrenfest+R predicts  $I_{\text{coh}} / I_{\text{tot}}$ quantitatively.
However, for strong incident fields, beyond the correct prediction of \textit{intensities}, one can wonder: will Ehrenfest+R also quantitatively capture the \textit{frequency dependence} of the scattered fields (which is not required in an Algorithm of Ehrenfest+R)? For example, can Ehrenfest+R recover the Mollow triplet \cite{Mollow1969} correctly? 
This challenging question will be studied in a future publication\cite{Chen2019Mollow}.

\section{Conclusion} \label{sec:conclusion}
In this paper, we have benchmarked the performance of five methods (the coupled Maxwell--Bloch equations, the classical OBE, Ehrenfest, Ehrenfest+R and CDT) for modeling light--matter interactions. When studying a TLS excited by an incident cw field, we find:
(i) Because Ehrenfest dynamics include only self-interaction for electronic relaxation, this method fails to correctly describe the electronic dynamics and optical signals  beyond the linear response regime; (ii)  The coupled Maxwell--Bloch equations and CDT fail to predict the correct optical signals even in the linear response regime, because  both methods effectively double-count  self-interaction: the self-interaction is accounted for through the spontaneous emission with  both the scattered field and an explicit dissipative term; see Eq. (\ref{eq:maxwell_bloch_rho}). As a consequence, these methods predict spectra with twice the correct FWHM in Fig. \ref{fig:1};
(iii)
Because both the classical OBE and the Ehrenfest+R approach carefully count self-interaction only once (but in different ways), these two methods describe both the electronic dynamics and the optical signals correctly in both the linear response regime and slightly beyond linear response. Moreover, because Ehrenfest+R preserves the self-interaction due to the scattered field and also enforces electronic relaxation due to vacuum fluctuations, this  approach
should be applicable for modeling light--matter interactions for a network of molecules, whereas for the OBE, one would need to carefully exclude the self-interaction due to the scattered field. Finally, by conserving energy, Ehrenfest+R can correctly distinguish coherent and total scattering over a wide range of light--matter couplings and results are in agreement with the quantum OBE.

The role of the total scattered field is essential for many collective phenomena, including, for example, resonant energy transfer and superradiance, and our laboratory is very excited to learn what new physics can be predicted with a powerful, new semiclassical approach to electrodynamics.

\begin{acknowledgement}
This material is based upon work supported by the U.S. Department of Energy, Office
of Science, Office of Basic Energy Sciences under Award Number DE-SC0019397. The authors thank Prof. Abraham Nitzan and Prof. Maxim Sukharev for stimulating discussions.

\end{acknowledgement}


\begin{appendices}
\section{Electronic Relaxation within Ehrenfest dynamics}\label{sec:Ehrho}

\renewcommand{\theequation}{A\arabic{equation}}
\renewcommand{\thefigure}{A\arabic{figure}}
One significant difference between Ehrenfest dynamics and the other methods studied in this paper is that the equation of motion for Ehrenfest dynamics has no explicit dissipative term; see Eq. (\ref{eq:Ehrenfest_rho}). One may ask, do Ehrenfest dynamics still recover electronic relaxation? Here, for completeness, we summarize the main points in Ref. \citenum{Chen2018Spontaneous}.

To understand electronic relaxation for Ehrenfest dynamics, we can split the total E-field into the incident field plus the scattered field: $\vE(\vR, t) = \vEin(\vR, t) + \vEscatt(\vR, t)$. While the incident field obeys $\vEin(\vR, t) = \vEin(\vR-ct, 0)$, we must explicitly propagate $\vEscatt(\vR, t)$ and solve  Maxwell's equations. In 1D, if we initialize  the TLS such that $\rho_{12}(0) = 0$ at time zero,  the solution for $\vEscatt$ is\cite{Li2018Spontaneous}
\begin{equation}\label{eq:Escatter_solution}
\vEscatt(x, t) = \frac{\omega_0}{c\epsilon_0} \int_{0}^{t} dt' \text{Im}\dot{\rho}_{12}(t')\int_{x-c(t-t')}^{x+c(t-t')}dx' \vxi(x')
\end{equation}
where $\vxi(x) = \mu_{12} \vdelta(x)$ if we assume a point-dipole, and $\dot{\rho}_{12}$ is the time derivative of $\rho_{12}$, so that $\hmu\cdot\vEscatt(\mathbf{0}, t) = \int dx \vxi(x)\cdot \vEscatt(x, t)\hsigma_z= \hbar\kFGR\text{Im}\rho_{12}\hsigma_z$, where $\hsigma_z = \bigl( \begin{smallmatrix}0 & 1\\ 1 & 0\end{smallmatrix}\bigr)$. By substituting Eq. (\ref{eq:Escatter_solution}) into Eq. (\ref{eq:Ehrenfest_rho}), one finally obtains
\begin{equation}
\begin{aligned}
\frac{d \hat{\rho} }{dt} &=  \Lrho{s} + \Lrho{E} \\
&= \Lrho{s} + \Lrho{E\textsubscript{in}} + \Lrho{Escatt}
\end{aligned}
\end{equation}
where $\Lrho{s}$ and $\Lrho{E\textsubscript{in}}$ are defined in Eqs. (\ref{eq:LsLEin}) and 
\begin{equation}\label{eq:Leh2}
\begin{aligned}
\mathcal{L}_{\text{Escatt}}[\hat{\rho}] &= \frac{i}{\hbar}\left[\hmu\cdot\vEscatt(\mathbf{0}, t), \hat{\rho}\right]  \\
&= 
\kFGR
\begin{pmatrix}
2\left[\text{Im}\rho_{12}\right]^2 & -i \text{Im}\rho_{12}\left(\rho_{11} - \rho_{22}\right) \\
-i \text{Im}\rho_{21}\left(\rho_{11} - \rho_{22}\right) & -2\left[\text{Im}\rho_{12}\right]^2
\end{pmatrix}
\end{aligned}
\end{equation}
Eq. (\ref{eq:Leh2}) clearly shows that the scattered field contributes to the electronic relaxation if $\text{Im}\rho_{12}\neq 0$. Furthermore, though not proven here, Eq. (\ref{eq:Leh2}) is valid in 3D (as well as 1D)\cite{Chen2018Spontaneous}. We will now analyze Eq. (\ref{eq:Leh2}) in the weak coupling limit.

\paragraph{Population relaxation rate} In the weak coupling limit,  $\omega_0 \gg \kFGR$, and we can define a time scale ($\tau$) $1/\omega_0 \ll \tau \ll 1/\kFGR$. With this time scale, one can assume $\rho_{12} \approx |\rho_{12}|e^{i\omega_0 t}$ so that $\left[\text{Im}\rho_{12}\right]^2 \approx |\rho_{12}|^2 \sin^2\left(\omega_0 t\right)$. We may then define an instantaneous decay rate $\kEh(t)$ for $\rho_{22}$, satisfying $\dot{\rho}_{22} = -\kEh(t)\rho_{22}$, where
\begin{equation}\label{eq:kEh-original}
\kEh(t) = \kFGR \frac{|\rho_{12}|^2}{\rho_{22}} \left  [2 \sin^2\left(\omega_0 t\right)\right]
\end{equation}
We call Eq. (\ref{eq:kEh-original}) the Ehrenfest decay rate.
If we average this rate over all relevant $\tau$, we find $\overline{\sin^2\left(\omega_0 t\right)} = 1/2$, and  one obtains Eq. (\ref{eq:kEh-coarse}).
For Ehrenfest dynamics, because no non-Hamiltonian term appears in the Liouville equation, purity is strictly conserved, so that $|\rho_{12}|^2 = \rho_{11} \rho_{22}$, and Eq. (\ref{eq:kEh-coarse}) reduces to Eq. (\ref{eq:kEh-kFGR}).


\paragraph{Dipole dephasing rate} When comparing the Ehrenfest effective dissipative term ($\Lrho{Escatt}$ in Eq. (\ref{eq:Leh2})) against the OBE ($\Lrho{SE}$ in Eq. (\ref{eq:Lrho_optical_bloch})), we find one interesting disagreement: for Ehrenfest dynamics, the off-diagonal dissipation is purely imaginary, but for the OBE, there is a real component. When averaged over a time $\tau$, the absolute values of these off-diagonal terms remain the same. However, as shown in Figs. \ref{fig:traj}b,d, this difference causes the electronic  dynamics for Ehrenfest dynamics to differ slightly as compared with the OBE when an off-resonant cw field excites the TLS. This difference can be explained as follows. On the one hand, for the classical OBE, if we consider only the effect of the dipole-dephasing for $\rho_{12}$, i.e., $\frac{d}{dt}\rho_{12} = -\gamma \rho_{12}$,  we can rewrite it as $\rho_{12}(t) = e^{-\gamma t}\rho_{12}(0)$ after an integral over time. 


On the other hand, for Ehrenfest dynamics, even though we still find $\frac{d}{dt}\rho_{12} = -\gamma \rho_{12}$,   $\gamma$ is now purely imaginary (see Eq. (\ref{eq:Leh2})). Hence, it is natural to consider the absolute value of $\rho_{12}$, which satisfies $\frac{d}{dt}|\rho_{12}| = -\gEh(t)|\rho_{12}|$, 
\begin{equation}\label{eq:gEh-original}
\gEh(t) = \kFGR \left(\rho_{11} - \rho_{22}\right) \sin^2\left(\omega_0 t\right)
\end{equation}
In a coarse-grained picture, $\gEh(t)$ reduces to Eq. (\ref{eq:gEh-coarse}).
When $\rho_{11} \rightarrow 1$, $\gEh$ agrees with the OBE ($\gamma_{\text{Bloch}} = \kFGR/2$) and Ehrenfest+R is not needed. Thus, the major difference between Ehrenfest+R and the OBE is just the phase of dipole dephasing rate $\gamma$, which can lead to a slight difference in $\rho_{12}$ at later times.



\section{The Detailed Procedure for the Ehrenfest+R Approach}\label{sec:EhR_procedure}
\renewcommand{\theequation}{B\arabic{equation}}
\renewcommand{\thefigure}{B\arabic{figure}}
\subsection{+R correction for the electronic dynamics}
According to Ehrenfest dynamics, self-interaction leads to some fraction of the true FGR rate of electronic relaxation ($\Lrho{E\textsubscript{scatt}}$ in Eq. (\ref{eq:Leh});  see also Eqs. (\ref{eq:kEh-coarse}-\ref{eq:gEh-coarse})). In order to correctly recover the full FGR decay rate, we include an additional dissipative ("+R") term ($\Lrho{R}$) on top of the normal Liouville equation in Eq. (\ref{eq:Ehrenfest_rho}). At every time step, we write 
\begin{equation}\label{eq:rho_ehR}
\hrho_{\text{Eh+R}}(t+dt) = \hrho_{\text{Eh}}(t+dt) + \mathcal{L}_{\text{R}}[\hrho_{\text{Eh}}(t+dt)]dt
\end{equation}
Here, $\hrho_{\text{Eh}}$ refers to the electronic density operator that is  propagated with Ehrenfest dynamics for one time step; $\Lrho{R}$ is defined as
\begin{equation}\label{eq:LR}
\Lrho{R}=
\begin{pmatrix}
\kR(t)\rho_{22} & -\gR(t)\rho_{12} \\
-\gR(t) \rho_{21} & -\kR(t)\rho_{22}
\end{pmatrix}
\end{equation}
where we define $\kR$ and $\gR$ to be the +R population relaxation and dipole dephasing rates, respectively:
\begin{subequations}\label{eq:kR_gR}
	\begin{align}
	\kR(t) &\equiv 2 \kFGR\left[1 - \frac{|\rho_{12}|^2}{\rho_{22}}\right] \text{Im}\left[\frac{\rho_{12}}{|\rho_{12}|} e^{i\phi^l}\right]^2 \label{eq:kR}\\
	\gR(t) &\equiv \frac{\kFGR}{2}\left(1 - \rho_{11} + \rho_{22}\right) \label{eq:gR}
	\end{align}
\end{subequations}
According to Eq. (\ref{eq:kR}), each trajectory $l$ experiences its own $\kR(t)$ with an arbitrary phase $\phi^l \in \left[0, 2\pi\right)$. Note that this phase does \textit{not} change during the simulation. From our point of view, introducing this stochastic element on top of Ehrenfest dynamics is entirely reasonable; and similar approaches have already been proposed in the context of nuclear-electronic dynamics\cite{Bedard-Hearn2005,Hack2001,Volobuev2000,Larsen2006}.

In a coarse-grained picture, averaging over the random phase $\phi^l$, one finds $\text{Im}\left[\frac{\rho_{12}}{|\rho_{12}|} e^{i\phi^l}\right]^2 = \frac{1}{2}$, so that
\begin{equation}
\overline{\kR}(t) = \kFGR \left(1 - \frac{|\rho_{12}|^2}{\rho_{22}}\right)
\end{equation}
Thus, the total emission (including the self-interaction of the scattered field in Ehrenfest dynamics ($\mathcal{L}_{\text{E\textsubscript{scatt}}}[\hat{\rho}]$ in Eq. (\ref{eq:Leh})) plus the +R quantum vacuum fluctuations pathway ($\mathcal{L}_{\text{R}}[\hat{\rho}]$ in Eq. (\ref{eq:LR}))) is identical to the total dissipation found in the OBE ($\Lrho{SE}$ in Eq. (\ref{eq:Lrho_optical_bloch})):
\begin{subequations}
	\begin{align}
	\overline{\kEh}(t) + \overline{\kR}(t) = \kFGR \\
	\overline{\gEh}(t) + \gR(t) = \frac{\kFGR}{2}
	\end{align}
\end{subequations}

\paragraph{+R Correction in the  Wave Function Picture}
One appealing quality of Ehrenfest dynamics is that the purity of the electronic subsystem does not change within a single trajectory, and  one can propagate Ehrenfest dynamics with a density operator $\hrho$ or a wave function $\vC$. The Ehrenfest+R approach is consistent with this structure, and can be implemented in the wave function picture as well:
\begin{equation}\label{eq:+R-C}
\vC_{\text{Eh+R}}(t+dt) = e^{i\hat{\Phi}[\gR]} \TkR \vC_{\text{Eh}}(t+dt)
\end{equation}
Here, $\vC_{\text{Eh}}$ is the quantum amplitude after one time step propagated according to Ehrenfest dynamics, and $\vC_{\text{Eh+R}}$ is the corresponding  quantum amplitude after the +R event.
The quantum transition operator $\TkR$ in Eq. (\ref{eq:+R-C}) is responsible for enforcing additional population relaxation, and changes $\vC$ to $\vC'$:
\begin{equation}
\TkR
\begin{pmatrix}
c_1 \\
c_2
\end{pmatrix}
=
\begin{pmatrix}
c_1' \\
c_2'
\end{pmatrix}
\end{equation}
For a TLS, the relation between $\vC'$ and $\vC$ is
\begin{subequations}\label{eq:c1c2+R}
	\begin{align}
	c_1' = \frac{c_1}{|c_1|}\sqrt{|c_1|^2 + \kR(t) |c_2|^2dt} \\
	c_2' = \frac{c_2}{|c_2|}\sqrt{|c_2|^2 - \kR(t) |c_2|^2dt} 
	\end{align}
\end{subequations}
where $\kR$ is defined in Eq. (\ref{eq:kR}).
When $c_1 = 0$ (and the electronic subsystem is in the excited state), $\kR = \kFGR$,  and $\frac{c_1}{|c_1|}$ and $\frac{c_2}{|c_2|}$ are not well defined. For a  practical implementation, when $c_1 = 0$ , we force $\frac{c_1}{|c_1|} = 1$ and we allow $\frac{c_2}{|c_2|} = e^{i\theta}$, where $\theta \in \left[0, 2\pi\right)$ is a random number. Note that $\theta$ and $\phi^l$ have no correlation.

In Eq. (\ref{eq:+R-C}), after invoking the quantum transition operator $\TkR$,   we perform a stochastic random phase operator $e^{i\hat{\Phi}[\gR]}$ to enforce the additional dipole dephasing:
\begin{eqnarray}\label{eq:T12}
e^{i\hat{\Phi}[\gR]} = 
\begin{cases}
\begin{pmatrix}
e^{i\Phi_0} & 0 \\
0 & 1
\end{pmatrix}, & \text{if RN $< \gR dt$ }\\
\hat{\mathds{1}}, & \text{otherwise}
\end{cases}
\end{eqnarray}
where $\hat{\mathds{1}}$ is the identity operator, $\Phi_0$  and $\text{RN}$ are independent random numbers with range $\Phi_0 \in \left[0, 2\pi\right)$ and $\text{RN} \in \left[0, 1\right)$. During the time interval $dt$, Eq. (\ref{eq:T12}) efficiently reduces the ensemble average coherence $\avg{c_1c_2^{\ast}}$ by an amount $\gR dt\avg{c_1c_2^{\ast}}$.

Note that, for Figs. \ref{fig:traj}-\ref{fig:2}, we have confirmed numerically that propagating the +R correction in the wave function picture (Eqs. \ref{eq:+R-C}-\ref{eq:T12}) yields the same results compared with the density matrix picture (Eqs. \ref{eq:rho_ehR}-\ref{eq:kR_gR}). However, for stronger incoming fields, the two methods will not always agree. For instance, the wave function picture that uses stochastic dephasing (Eq. (\ref{eq:T12})) can sucessfully predict a Mollow triplet, while the density matrix picture fails to do so; see Ref. \citenum{Chen2019Mollow} for more details. Hence, we recommend always implementing Ehrenfest+R with the wave function picture.

\subsection{Rescaling the EM Field}
After enforcing the +R correction for the electronic subsystem, in order to conserve energy, one needs to rescale the classical EM field at each time step $dt$ by giving energy $\UR$ to the EM field:
\begin{equation}\label{eq:dURdt-+R}
\frac{d}{dt} \UR = \hbar \omega_0 \kR(t) \rho_{22} 
\end{equation}
In practice, we rescale the EM field after every time step $dt$ by
\begin{subequations}\label{eq:Ehrenfest+R-rescaleEM1}
	\begin{align}
	\vE_{\text{Eh+R}}^l &= \vE_{\text{Eh}}^l + \alpha^l \delta\vE_\text{R} \\
	\vB_{\text{Eh+R}}^l &= \vB_{\text{Eh}}^l + \beta^l \delta\vB_\text{R} 
	\end{align}
\end{subequations}
where $\vE^l_{\text{Eh}}$ denotes the E-field for trajectory $l$ with no field rescaling and $\vE_{\text{Eh+R}}^l$ denotes the E-field after field rescaling;
the rescaling functions  $\delta\vE_\text{R}$ and $\delta\vB_\text{R}$ are
chosen according to polarization density and these fields should not self-interfere with the TLS or otherwise
influence the propagation of $\hat{\rho}$ (because the addtion of $\Lrho{R}$ already leads to the correct spontaneous decay rate). 
In 1D, for the polarization profile defined in Eq. (\ref{eq:vxi}), these rescaling functions are defined as
\begin{subequations}\label{eq:Ehrenfest+R-rescaleEM2}
	\begin{align}
	\delta\vE_\text{R}(x) &= - \frac{\mu_{12}}{\sqrt{2\pi}\sigma^5}x^2e^{-\frac{x^2}{2\sigma^2}}\ve_z \\
	\delta\vB_\text{R}(x) &=  \frac{\mu_{12}}{3\sqrt{2\pi}\sigma^5}x^3e^{-\frac{x^2}{2\sigma^2}}\ve_y
	\end{align}
\end{subequations}
More generally, in 3D, $\delta\vE_\text{R} = \vnabla\times\vP\times\vP$ and $\delta\vB_\text{R} = \vnabla\times \vP$.\cite{Mukamel1999,VanKranendonk1977}
The parameters $\alpha^l$ and $\beta^l$ are defined to conserve the total energy:
\begin{subequations}\label{eq:alpha_beta}
	\begin{align}
	\alpha^l &= \text{sgn}\left(\text{Im}\left[\rho_{12}e^{i\phi^l}\right]\right) \sqrt{\frac{c}{\Lambda}\frac{\dot{U}_\text{R}}{\epsilon_0\int dv |\delta\vE_\text{R}|^2}} dt\\
	\beta^l &= \text{sgn}\left(\text{Im}\left[\rho_{12}e^{i\phi^l}\right]\right) \sqrt{\frac{c}{\Lambda}\frac{\mu_0 \dot{U}_\text{R}}{\int dv |\delta\vB_\text{R}|^2}} dt
	\end{align}
\end{subequations} 
Here, $\dot{U}_\text{R}$ is short for $\frac{d}{dt}\UR$, and $\UR$ is defined in Eq. (\ref{eq:dURdt-+R});   $\Lambda$ is the self-interference length, which is defined as
\begin{equation}\label{eq:Lambda}
\Lambda = \frac{2\pi^2 \left| \delta\widetilde{\vE}_{\text{R}}(0)\right|^2}{\int d\vR \left|\delta\vE_{\text{R}}\right|^2}
+  \frac{2\pi^2 \left| \delta\widetilde{\vB}_{\text{R}}(0)\right|^2}{\int d\vR \left|\delta\vB_{\text{R}}\right|^2}
\end{equation}
$\delta\widetilde{\vE}_{\text{R}}$ and $\delta\widetilde{\vB}_{\text{R}}$ are the Fourier components of the rescaling fields $\delta\vE_{\text{R}}$ and $\delta\vB_{\text{R}}$:
\begin{subequations}
	\begin{align}
	\delta\vE_{\text{R}}(\vR)  = \int d\mathbf{k} \ \delta\widetilde{\vE}_{\text{R}}(\mathbf{k}) e^{i\mathbf{k}\cdot \vR}\\
	\delta\vB_{\text{R}}(\vR)  = \int d\mathbf{k} \ \delta\widetilde{\vB}_{\text{R}}(\mathbf{k})e^{i\mathbf{k}\cdot \vR}
	\end{align}
\end{subequations} 
For the polarization profile in Eq. (\ref{eq:vxi}),  we find $\Lambda = \frac{4\sqrt{\pi}}{3}\sigma = 2.363\sigma$.

\section{Optical Polarization for Each Method}\label{sec:OpticalPolarization}

\renewcommand{\theequation}{C\arabic{equation}}
\renewcommand{\thefigure}{C\arabic{figure}}

\subsection{The optical Bloch equation}
For the OBE, one  calculates the effective optical polarization, $\vPtot(t)$, according to  the mean-field approximation:
\begin{equation}\label{eq:P-mean-field-approx}
\vPtot(t) = \tr{\hat{\rho}(t)\hmu}
\end{equation}
Here, one needs to be careful about notation. $\vPtot$ denotes the total optical polarization, which is the integral over the polarization density operator, $\hP(\vR)$ in Eq. (\ref{eq:P}): $\vPtot = \int d\vR \tr{\hrho \hP(\vR)}$.

By taking the second-order time derivative of Eq. (\ref{eq:P-mean-field-approx}), using $\vPtot = 2\text{Re}\rho_{12}\mu_{12}\ve_z$, calculating $\frac{d}{dt}\text{Re}\rho_{12}$ and $\frac{d}{dt}\text{Im}\rho_{12}$ and further applying Eqs. (\ref{eq:hmu}) and (\ref{eq:optical_bloch}), the equation of motion for $\vPtot(t)$ can be expressed as 
\begin{equation}\label{eq:ddPddt_optical}
\ddot{\vPtot}(t) + \kFGR \dot{\vPtot}(t) + \omega_0^2 \vPtot(t) = \epsilon_0\omega_p^2 W_{12}(t)  \vE_{\text{in}}(t)
\end{equation}
Here, we define $\omega_p \equiv \sqrt{2\mu_{12}^2\omega_0 /\epsilon_0\hbar}$ to be the plasmon frequency,  $W_{12} \equiv \rho_{11} - \rho_{22}$, $\vEin(t)$ is short for $\vEin(\mathbf{0}, t)$. 

Eq. (\ref{eq:ddPddt_optical}) is an anharmonic oscillator picture for optical polarization\cite{Mukamel1999}. Given an initial condition for $\vPtot$, according the classical OBE, one  evolves Eq. (\ref{eq:ddPddt_optical}) coupled with Maxwell's equations in Eqs. (\ref{eq:maxwell_bloch_EM}) to obtain the optical signals for a TLS.

\subsection{The coupled Maxwell--Bloch equations}
Following the procedure for the OBE, one obtains a very similar equation of motion  for optical polarization $\vPtot$ in the case of the coupled Maxwell--Bloch equations:
\begin{equation}\label{eq:ddPddt_coupled}
\ddot{\vPtot}(t) + \kFGR \dot{\vPtot}(t) + \omega_0^2 \vPtot(t) = \epsilon_0 \omega_p^2 W_{12}(t) \vE(t)
\end{equation}
Here, $\omega_p$ and $W_{12}$ are defined the same as in Eq. (\ref{eq:ddPddt_optical}).
Comparing Eq. (\ref{eq:ddPddt_coupled}) to Eq. (\ref{eq:ddPddt_optical}), because $\vE = \vEin + \vEscatt$, we see that  the optical polarization can be significantly different if the scattered field is not negligible.

\subsection{Ehrenfest dynamics}
For Ehrenfest dynamics, from Eq. (\ref{eq:Ehrenfest}), the equation of motion for the optical polarization is easy to derive:
\begin{equation}\label{eq:ddPddt-Ehrenfest}
\ddot{\vPtot}(t) + \omega_0^2 \vPtot(t) = \epsilon_0 \omega_p^2 W_{12}(t)\vE(t)
\end{equation}
Compared to the other equations of motion for $\vPtot$ in Eqs. (\ref{eq:ddPddt_optical}) and (\ref{eq:ddPddt_coupled}), the major difference is that Eq. (\ref{eq:ddPddt-Ehrenfest}) has no explicit relaxation term ($\kFGR \dot{\vPtot}$). However, the lack of such a relaxation term does not imply that the anharmonic oscillator will not be damped to zero at long times. In fact, since $\vPtot = \tr{\hat{\rho}\hmu}$, and $\hat{\rho}$ is relaxed by the scattered field $\vEscatt$ (see the discussion in Sec. \ref{sec:Eh-dissipation}), $\vPtot$ will eventually be damped to zero  as long as the TLS is not initiated exactly in the excited state. More explicitly, if we separate the electric field as $\vE = \vEscatt + \vEin$, 
and use the fact that  $\vmu\cdot\vEscatt = \hbar\kFGR\text{Im}\rho_{12} = -\frac{\hbar\kFGR}{\omega_0}\frac{d\text{Re}\rho_{12}}{dt}$, where we denote $\vmu = \mu_{12} \ve_z$,	
Eq. (\ref{eq:ddPddt-Ehrenfest}) can be rewritten as
\begin{equation}\label{eq:ddPddt-Ehrenfest-2}
\ddot{\vPtot}(t) + W_{12}(t)\kFGR\dot{\vPtot} + \omega_0^2 \vPtot(t) = \epsilon_0 \omega_p^2 W_{12}(t)\vEin(t)
\end{equation}
Here,  the effective relaxation term $W_{12}(t)\kFGR\dot{\vPtot}$ causes a population-dependent damping for $\vPtot$.

\subsection{The Ehrenfest+R approach}
For Ehrenfest+R, the equation of motion for optical polarization reads
\begin{equation}\label{eq:ddPddt_EhR}
\ddot{\vPtot}(t) + 2\gR(t) \dot{\vPtot}(t) + \left[\omega_0^2 + \dot{\gamma}_{\text{R}}(t)  + \gamma_{\text{R}}^2(t)\right] \vPtot(t) = \epsilon_0 \omega_p^2 W_{12}(t)\vE(t)
\end{equation}
where $\gR$ is defined in Eq. (\ref{eq:kR_gR}). 
To derive Eq. (\ref{eq:ddPddt_EhR}), we simply take advantage of $\frac{d}{dt}\hrho = -\frac{i}{\hbar}\left[\hH_{sc}, \hrho\right] - \Lrho{R}$, where $\hH_{sc}$ and $\Lrho{R}$ are defined in Eqs. (\ref{eq:Ehrenfest_Hsc}) and (\ref{eq:LR}), and find
\begin{subequations}
	\begin{align}
	\frac{d \text{Re}\rho_{12}}{dt} &= -\omega_0\text{Im}\rho_{12} - \gR(t)\text{Re}\rho_{12}\label{eq:dRerho12dt} \\
	\frac{d \text{Im}\rho_{12}}{dt} &= \omega_0\text{Re}\rho_{12} - W_{12}\vmu\cdot \vE - \gR(t)\text{Im}\rho_{12} \label{eq:dImrho12dt}
	\end{align}
\end{subequations}
 By taking the time derivative of Eq. (\ref{eq:dRerho12dt}) and applying Eq. (\ref{eq:dImrho12dt}), we obtain
\begin{equation}\label{eq:d2Rerho12dt}
\frac{d^2\text{Re}\rho_{12}}{dt^2} = -\omega_0\left[\omega_0\text{Re}\rho_{12} - W_{12}\vmu\cdot \vE - \gR(t)\text{Im}\rho_{12} \right] - \frac{d\gR(t)}{dt} \text{Re}\rho_{12} - \gR(t)\frac{d\text{Re}\rho_{12}}{dt}
\end{equation}
Note that, if we set $\gR(t) = 0$, we recover Eq. (\ref{eq:ddPddt-Ehrenfest}). Nevertheless, if $\gR(t) \neq 0$, note also that 
Eq. (\ref{eq:d2Rerho12dt}) still contains $\text{Im}\rho_{12}$. To eliminate this term, we can rewrite Eq. (\ref{eq:dRerho12dt}) as $\text{Im}\rho_{12} = \frac{1}{\omega_0}\left[\gR(t)\text{Re}\rho_{12} - \frac{d\text{Re}\rho_{12}}{dt}\right]$. By substituting this identity into Eq. (\ref{eq:d2Rerho12dt}), we finally derive Eq. (\ref{eq:ddPddt_EhR}).

In Eq. (\ref{eq:ddPddt_EhR}),
by introducing the relaxation term $2\gR(t) \dot{\vPtot}(t)$ where $\gR(t)$ depends on the electronic state, one recovers the FGR rate correctly as compared with Ehrenfest dynamics. One interesting feature of Eq. (\ref{eq:ddPddt_EhR}) [as compared with Eqs. (\ref{eq:ddPddt_optical}-\ref{eq:ddPddt-Ehrenfest})] is that the intrinsic frequency is no longer $\omega_0$ but rather $\sqrt{\omega_0^2 + \dot{\gamma}_{\text{R}}(t) + \gamma_{\text{R}}^2(t)}$. This frequency renormalization will be studied in a future publication.

\subsection{CDT}
The susceptibility $\chi(\omega)$, which is derived in Appendix \ref{sec:Chi} for a TLS, plays a significant role in classical electrodynamics.
If we take the Fourier transform of the definition of $\chi(\omega)$ in Eqs. (\ref{eq:PE-identity}-\ref{eq:chi_L}), we find that the equation of motion for the optical polarization reads
\begin{equation}\label{eq:ddPddt_fdtd}
\ddot{\vPtot}(t) + \kFGR \dot{\vPtot}(t) + \omega_0^2 \vPtot(t) = \epsilon_0 \omega_p^2  \vE(t)
\end{equation}
Eq. (\ref{eq:ddPddt_fdtd}) is very similar to Eq. (\ref{eq:ddPddt_coupled}) in the weak-excitation limit ($W_{12} \rightarrow 1$), indicating that a standard CDT treatment also double-counts self-interaction, just as do the coupled Maxwell--Bloch equations. This double-counting originates from the inconsistency between the derivation of $\chi(\omega)$\cite{Boyd2008} (where we assume $\vPtot(\omega) = \epsilon_0 \chi(\omega) \vEin(\omega)$), and the numerical implementation of CDT (where people frequently take $\vPtot(\omega) = \epsilon_0 \chi(\omega) \vE(\omega)$ for a practical simulation); see Sec. \ref{sec:Chi}.

\section{Energy Conservation for Each Method}\label{sec:EnergyConserve}
\renewcommand{\theequation}{D\arabic{equation}}
\renewcommand{\thefigure}{D\arabic{figure}}

We define the energy of the quantum subsystem as
\begin{equation}\label{eq:Us}
\Us(t) = \tr{\hat{\rho}(t) \hH_s}
\end{equation}
For a classical EM field, the energy is defined as
\begin{equation}\label{eq:UEM}
\UEM(t) = \frac{1}{2}\int d\vR  \left(\epsilon_0|\vE(\vR, t)|^2 + \frac{1}{\mu_0}|\vB(\vR, t)|^2\right)
\end{equation}
For semiclassical approaches, the total energy is expressed as $\Ut= \Us+ \UEM$. 

\subsection{The optical Bloch equation}
For the OBE, similar to Ehrenfest dynamics (see Eq. (\ref{eq:Ehrenfest-Lform})), we can express Eq. (\ref{eq:optical_bloch}) as
\begin{equation}\label{eq:optical-Bloch-Lform}
\frac{d}{dt}\hat{\rho} = \Lrho{s}  + \Lrho{E\textsubscript{in}} + \Lrho{SE}
\end{equation}
where $\Lrho{s}$ and $\Lrho{E\textsubscript{in}}$ are defined in Eqs. (\ref{eq:LsLEin}) and $\Lrho{SE}$ is defined in Eq. (\ref{eq:Lrho_optical_bloch}).
Substituting Eq. (\ref{eq:optical-Bloch-Lform}) into Eq. (\ref{eq:Us}), the energy loss rate for the TLS reads 
\begin{equation}\label{eq:dUsdt-optical}
\frac{d}{dt} \Us =  \tr{\Lrho{E\textsubscript{in}}\hH_s} + \tr{\Lrho{SE}\hH_s} 
\end{equation}
By taking the time derivative of Eq. (\ref{eq:UEM}), and applying Maxwell's equations as defined in Eq. (\ref{eq:Ehrenfest}), the energy gain rate of the EM field is
\begin{equation}\label{eq:dUEMdt-optical}
\frac{d}{dt} \UEM =  \oint_{\varSigma} d\mathbf{s} \ \vE \times \vB - \int d\vR \  \vE \cdot \vJ
\end{equation}
Here, $\varSigma$ denotes the sphere in real space over which we integrate. If we integrate over a large sphere, the first term on the RHS in Eq. (\ref{eq:dUEMdt-optical}) vanishes. 

However, because one cannot find an exact cancellation between Eqs. (\ref{eq:dUsdt-optical}) and (\ref{eq:dUEMdt-optical}), the total energy of the classical OBE is not conserved. That being said, if the EM dynamics are propagated with the quantum EM field instead of classical EM field, the quantum OBE should conserve energy, provided we use the full quantum energy of the EM field,
\begin{equation}\label{eq:UEM-QED}
\begin{aligned}
\UEM'(t) = \frac{1}{2}\int d\vR  \left(\epsilon_0\avg{\hE(\vR, t)^2} + \frac{1}{\mu_0}\avg{\hB(\vR, t)^2}\right) 
\end{aligned}
\end{equation}
For more details about the quantum OBE and Eq. (\ref{eq:UEM-QED}), see Ref. \citenum{Cohen-Tannoudji1998}.

\subsection{The coupled Maxwell--Bloch equations}
Just as for the classical OBE, according to
the coupled Maxwell--Bloch equations (Eq. (\ref{eq:maxwell_bloch_rho})), 
the energy loss rate for a TLS reads
\begin{equation}
\frac{d}{dt} \Us = \tr{\Lrho{E} \hH_s} +  \tr{\Lrho{SE} \hH_s} 
\end{equation}
The energy gain for the EM field is the same as Eq. (\ref{eq:dUEMdt-optical}). After some straightforward algebra, we find the total energy obeys
\begin{equation}\label{eq:dUtdt-maxwell-bloch}
\frac{d}{dt} \Ut =   \tr{\Lrho{SE} \left(\hH_s - \hmu\cdot\vE(\mathbf{0}, t) \right)}  \neq 0
\end{equation}
In most situations, the magnitude of the coupling $|\hmu\cdot\vE(\mathbf{0}, t)|$ is much less than  $\hbar\omega_0$. Under such conditions, we can further simplify Eq. (\ref{eq:dUtdt-maxwell-bloch}) to read
\begin{equation}\label{eq:dUtdt-maxwell-bloch-2}
\frac{d}{dt} \Ut \approx   \tr{\Lrho{SE} \hH_s}  = -\hbar\omega_0 \kFGR \rho_{22} < 0
\end{equation}
Eqs. (\ref{eq:dUtdt-maxwell-bloch}-\ref{eq:dUtdt-maxwell-bloch-2}) show that energy is not conserved for the coupled Maxwell--Bloch equations and the total energy is continuously decreasing with a speed of $\hbar\omega_0\kFGR\rho_{22}$. This failure cannot be corrected by propagating quantum fields and using QED (as in Eq. (\ref{eq:UEM-QED})). The fundamental problem is not a quantum-classical mismatch but rather a double-counting of self-interaction.

\subsection{Ehrenfest dynamics}
As discussed above, for Ehrenfest dynamics, the energy loss rate for a TLS is expressed as
\begin{equation}
\frac{d}{dt} \Us = \tr{\Lrho{E}\hH_s}
\end{equation}
where $\Lrho{E}$ is defined as
\begin{equation}
\Lrho{E} =   \frac{i}{\hbar}\left[\hmu\cdot\vE(\mathbf{0}, t), \hrho\right]
\end{equation}
Since the energy gain rate for the EM field is defined just as in Eq. (\ref{eq:dUEMdt-optical}), the time derivative of the total energy can be expressed as
\begin{equation}
\begin{aligned}
\frac{d}{dt} \Ut &= \tr{\Lrho{E}\hH_s} - \int d\vR \ \vE \cdot \vJ \\
&= \tr{\Lrho{E}\hH_s} - \vE(\mathbf{0}, t) \cdot \tr{\Lrho{s} \hmu}\\
&= 0
\end{aligned}
\end{equation}
where we apply $\vJ = \tr{\dot{\hrho}\hP} = \tr{\dot{\hrho}\hmu}\vdelta(\vR)$ and $\dot{\hrho} = \Lrho{s} + \Lrho{E}$ for Ehrenfest dynamics. Here, $\Lrho{s}$ is defined in Eq. (\ref{eq:LsLEin}).
Thus, energy is conserved for Ehrenfest dynamics if we integrate over all space (where E and B fields vanish at the boundary). 

\subsection{The Ehrenfest+R approach}\label{sec:Ehrenfest+R-energy}
Ehrenfest+R is designed to yield energy conservation  provided one averages over trajectories, the total average Ehrenfest+R energy reads
\begin{equation}
\Ut = \tr{\hrho \hH_s} + \frac{1}{2}\int d\vR  \left(\epsilon_0\avg{|\vE_{\text{Eh+R}}|^2}_l + \frac{1}{\mu_0}\avg{|\vB_{\text{Eh+R}}|^2}_l\right)
\end{equation}
where $\avg{\cdots}_l$ denotes an ensemble average over trajectories indexed by $l \in \{1, 2,\cdots, N\}$. Here, $N$ is the total number of trajectories.
Because Ehrenfest dynamics alone conserve energy, we need  to  show only that the energy loss for the quantum subsystem due to the +R correction  balances the energy gain for the EM field due to the  rescaling of fields at every time step $dt$. This can be proved by induction: (i) suppose that, at time $t$, the total energy is conserved; (ii) at $t+dt$, $d\UR(t+dt) =  \hbar \omega_0 \kR(t+dt) \rho_{22}dt$ needs to be dissipated to the classical EM field; see Eq. (\ref{eq:dURdt-+R}). After the rescaling of EM field by Eq. (\ref{eq:Ehrenfest+R-rescaleEM1}-\ref{eq:alpha_beta}), the energy increment for the EM field has three components: (1) the squared norm of the added EM field at $t+dt$, (2) the product of the added EM field at $t+dt$ with the EM field generated by Ehrenfest dynamics, (3) and the product of the added EM field at $t+dt$ with the previously added EM fields ($\tau \leq t$). The overall increase in energy for the EM fields at $t+dt$ reads: 
	\begin{equation}\label{eq:dUR-conserve+R}
	\begin{aligned}
	d U_{\text{EM}}(t+dt) = & \ \frac{1}{2}\int d\vR  \left(\epsilon_0|\alpha^l_{t+dt}\delta\vE_{\text{R}}|^2 + \frac{1}{\mu_0}|\beta^l_{t+dt}\delta \vB_{\text{R}}|^2 \right)  \\
	+&  \int d\vR  \left(\epsilon_0\alpha^l_{t+dt}\delta\vE_{\text{R}}\cdot \vE_{\text{Eh}} + \frac{1}{\mu_0}\beta^l_{t+dt}\delta \vB_{\text{R}} \cdot \vE_{\text{Eh}}\right) \\
	+ & \sum_{\tau\leq t}\int d\vR  \left(\epsilon_0\alpha^l_{t+dt}\delta\vE_{\text{R}}\cdot \alpha^l_{\tau }\delta\vE_{\text{R}} + \frac{1}{\mu_0}\beta^l_{t+dt}\delta \vB_{\text{R}}
	\cdot \beta^l_{\tau}\delta \vB_{\text{R}} \right)
	\end{aligned}
	\end{equation}
We now average over the ensemble of trajectories: $\avg{d \UEM(t+dt)}_l$, the second integral in Eq. (\ref{eq:dUR-conserve+R}) vanishes because $\avg{\alpha^l}_l = \avg{\beta^l}_l = 0$ by definition. The third integral represents the self-interference between the current augmented field and the history of the past augmented EM fields, which is not zero. By carefully defining the prefactors $\alpha$ and $\beta$ (see Eqs. (\ref{eq:alpha_beta}-\ref{eq:Lambda})) to reflect how long an E or B field remains in the domain of self interaction,  the energy increment for the EM fields $d\UEM(t+dt)$ can be exactly balanced by $d\UR(t+dt)$; see Ref. \cite{Chen2018Spontaneous}. 

\section{Deriving the Dielectric Function for a TLS}\label{sec:Chi}
\renewcommand{\theequation}{E\arabic{equation}}
\renewcommand{\thefigure}{E\arabic{figure}}
An accurate expression of $\chi(\omega)$ (or $\epsilon(\omega)$) plays a key role for classical electrodynamics. The most standard approach to obtain $\chi(\omega)$ for a TLS is to use the OBE. Although this approach has been discussed in many textbooks\cite{Mukamel1999,Boyd2008}, to facilitate our discussion, we briefly outline the relevant procedure here. 

For a TLS excited by an incident cw field of frequency $\omega$, because $W_{12}(t) = \rho_{22} - \rho_{11}$ is a constant in the steady state, after a Fourier transform of Eq. (\ref{eq:ddPddt_optical}), the OBE yields
\begin{equation}\label{eq:dPdt_original_chi}
\begin{aligned}
-\omega^2 \vPtot(\omega) + i \omega \kFGR \vPtot(\omega) + \omega_0^2 \vPtot(\omega) = \epsilon_0 \omega_p^2 W_{12}^{ss} \vEin(\omega)
\end{aligned}
\end{equation}
where the superscript $^{ss}$ represents the steady-state solution.
Then, by using the identity 
\begin{equation}\label{eq:PchiE}
\vPtot(\omega) = \epsilon_0 \chi(\omega)\vEin(\omega)
\end{equation}
one obtains
\begin{equation}\label{eq:chi_ss}
\chi(\omega) =W_{12}^{ss} \frac{\omega_p^2}{\omega_0^2 - \omega^2 + i\omega \kFGR}
\end{equation}
Now the steady-state $W_{12}^{ss}$ can be calculated using the pseudospin form of the OBE\cite{Cohen-Tannoudji1998} after the rotating wave approximation. More precisely, if we express the OBE in terms of the variables $\hat{S}_{+}(t)$, $\hat{S}_{-}(t)$, and $\hat{S}_z(t)$, where $\hat{S}_z = \frac{1}{2}\left(\ket{e}\bra{e} - \ket{g}\bra{g}\right)$ and $\hat{S}_{+}$ and $\hat{S}_{-}$ are defined in Sec. \ref{sec:methods}, and then set the time derivatives of these variables to be zero, the pseudospin form of the OBE yields
\begin{equation}\label{eq:W_12_ss}
\begin{aligned}
W_{12}^{ss} &\equiv \rho_{11}^{ss} - \rho_{22}^{ss} \\
&= \frac{\kFGR^2 + 4(\omega - \omega_0)^2}{\kFGR^2 + 4(\omega - \omega_0)^2 + 2 |\mu_{12}|^2 |E_0|^2/\hbar^2}
\end{aligned}
\end{equation}
In the linear response regime ($E_0 \rightarrow 0$),  $W_{12}^{ss} \rightarrow 1$, and the dielectric function defined  in Eq. (\ref{eq:chi_ss}) can be reduced to
\begin{equation}\label{eq:E5}
\chi^{\text{L}}(\omega) \approx \frac{\omega_p^2}{\omega_0^2 - \omega^2 - i\omega \kFGR}
\end{equation}
which corresponds to a Lorentz medium.
Beyond linear response, after a Taylor expansion of Eq. (\ref{eq:W_12_ss}) as a function of $|E_0|^2$, the corresponding dielectric function becomes
\begin{equation}\label{eq:chiNL_series}
\chi^{\text{NL}}(\omega) \approx \chi^{\text{L}}(\omega) \left[1 - \frac{1}{1 + 4(\omega - \omega_0)^2/\kFGR^2 }  \frac{|E_0|^2}{|E_s|^2}   + \cdots \right]
\end{equation}
where we define $|E_s|^2 \equiv \hbar^2 \kFGR^2 / 2 |\mu_{12}|^2$, and expand to the lowest nonlinear order (third order). One interesting issue for Eq. (\ref{eq:chiNL_series}) is that this series converges only when $|E_0|/|E_s| < 1$. For $|E_0|/|E_s|$ not very much smaller than one, higher order nonlinear terms are required to enforce convergence; otherwise an CDT simulation becomes unstable. 

Note that in a regular CDT calculation, one simulates the total scattered field, so one calculates $\vPtot$ by $\vPtot = \epsilon_0\chi \vE$; this working equation  conflicts with the definition of $\chi$ in Eq. (\ref{eq:PchiE}) because $\vE \neq \vE_{\text{in}}$. This conflict causes a double-counting of self-interaction, in a  manner similar to the case of the coupled Maxwell--Bloch equations.

\section{The FDTD technique}\label{sec:fdtd}
\renewcommand{\theequation}{F\arabic{equation}}
\renewcommand{\thefigure}{F\arabic{figure}}
For the FDTD calculation, we propagate $\vH$, $\vD$ and $\vPtot$ in the time domain. Thereafter, we evaluate $\vE = \frac{1}{\epsilon_0}(\vD - \vPtot)$. To numerically propagate $\vPtot$ in the time domain,  one needs to take the inverse Fourier transform of $\vPtot(\omega) = \epsilon_0 \chi(\omega)\vE(\omega)$. For the case of a Lorentz medium defined by Eq. (\ref{eq:E5}), one finds the following  equation of motion for $\vPtot$:
\begin{equation}\label{eq:hehe}
\ddot{\vPtot}(t) + \kFGR \dot{\vPtot}(t) + \omega_0^2 \vPtot(t) = \epsilon_0 \omega_p^2  \vE(t)
\end{equation}
In Eq. (\ref{eq:hehe}), the lowest-order discretizations of time derivatives for $\vPtot$ are: $\dot{\vPtot}(n\Delta t) \approx \frac{\vPtot\left((n + 1)\Delta t\right) -  \vPtot\left((n - 1)\Delta t\right)}{2\Delta t}$ and  $\ddot{\vPtot}(n\Delta t) \approx \frac{\vPtot\left((n + 1)\Delta t\right) - 2\vPtot\left(n\Delta t\right) + \vPtot\left((n - 1)\Delta t\right)}{\Delta t^2}$, where $n$ is the index of time step and $\Delta t$ is the time interval between the neighboring time steps. Substituting these identities into Eq. (\ref{eq:hehe}) and reorganize the equation, one finally obtains
\begin{equation}\label{eq:ddPddt_discrete}
P^{n+1}_z(j) = \frac{1}{\frac{1}{\Delta t^2} + \frac{\kFGR}{2\Delta t}} \left[\epsilon_0\omega_p^2 E^{n}_z(j) - \left(-\frac{2}{\Delta t^2} + \omega_0^2\right)P^{n}_z(j) - \left(\frac{1}{\Delta t^2} - \frac{\kFGR}{2\Delta t}\right)P^{n-1}_z(j)\right]
\end{equation} 
where we assume $\vPtot$ and $\vE$ are oriented along the $z$-axis and $\vB$ is oriented along the $y$-axis; the superscript $n$ denotes the $n$-th time step and the  index $j$ denotes the $j$-th grid in space. According to Eq. (\ref{eq:ddPddt_discrete}), in order to propagate $P_z$ numerically to time step $n+1$, one needs to save the data for $P_z$ in the previous two time steps ($n-1$ and $n$) and update them at every time step. One can numerically propagate the  Maxwell's equations defined in Eq. (\ref{eq:Maxwell-classical}) by propagating Eq. (\ref{eq:ddPddt_discrete}) with
	\begin{subequations}\label{eq:FDTD_Maxwell}
		\begin{align}
		D_z^{n+1/2}(j) &= D_z^{n-1/2}(j) + \frac{\Delta t}{\Delta x} \left[H_y^n\left(j + \frac{1}{2}\right) - H_y^n\left(j - \frac{1}{2}\right)\right]  \\
		  E_z^{n+1/2}(j)  &= \frac{1}{\epsilon_0}\left[D_z^{n+1/2}(j) - P_z^{n+1/2}(j)\right]\\
		H_y^{n+1}\left(j+\frac{1}{2}\right) &= H_y^{n}\left(j+\frac{1}{2}\right)  - \frac{\Delta t}{\mu_0\Delta x}  \left[
		E_z^{n+1/2}\left(j +1\right) - E_z^{n-1/2}\left(j\right)
		\right] 
		\end{align}
	\end{subequations}
where $\Delta x$ is the spatial separation between the neighboring grids. The coefficient $1/2$ in  Eq. (\ref{eq:FDTD_Maxwell}) indicates that a staggered grid for $D_z$ and $H_y$ is used, which is  known as the Yee cell\cite{Yee1966}.
See Ref. \citenum{Taflove2005} for a far more detailed account of FDTD.

\end{appendices}

\bibliographystyle{achemso}

\begin{mcitethebibliography}{77}
\bibitem[foo()]{footnote1}
	Interestingly, near the ground state, we can still find a small difference
	between Ehrenfest dynamics and the optical Bloch equation. As discussed in
	Appendix \ref{sec:Ehrho}, when $\rho_{22}\rightarrow 0$, after a
	course-grained average, the diagonal elements of
	$\Lrho{E\textsubscript{scatt}}$ are exactly $\Lrho{SE}$; however, the
	off-diagonal elements of $\Lrho{E\textsubscript{scatt}}$ are purely imaginary
	(see Eq. (\ref{eq:Leh2})), while the off-diagonal elements of $\Lrho{SE}$ are
	not purely imaginary --- and yet the absolute values of both approaches are
	equal. This slight difference can cause minor difference in electronic
	dynamics; see Fig. \ref{fig:traj}b,d\relax
	\mciteBstWouldAddEndPuncttrue
	\mciteSetBstMidEndSepPunct{\mcitedefaultmidpunct}
	{\mcitedefaultendpunct}{\mcitedefaultseppunct}\relax
	\EndOfBibitem
\end{mcitethebibliography}

\providecommand{\latin}[1]{#1}
\makeatletter
\providecommand{\doi}
{\begingroup\let\do\@makeother\dospecials
	\catcode`\{=1 \catcode`\}=2 \doi@aux}
\providecommand{\doi@aux}[1]{\endgroup\texttt{#1}}
\makeatother
\providecommand*\mcitethebibliography{\thebibliography}
\csname @ifundefined\endcsname{endmcitethebibliography}
{\let\endmcitethebibliography\endthebibliography}{}

\end{document}